\documentclass[11pt,onecolumn]{article}
\usepackage{graphics,graphicx,epsfig,amsmath,color}
\usepackage[T2A]{fontenc} 
\usepackage[left=2cm,right=2cm,top=2cm,bottom=2cm,bindingoffset=0cm]{geometry}
\usepackage{amsmath, amsthm, amssymb}
\usepackage{graphicx}
\usepackage[utf8]{inputenc}
\usepackage[english]{babel} 
\usepackage{cite}
\usepackage{enumitem}
\usepackage{float}
\usepackage{color}
\usepackage{xcolor}
\usepackage{booktabs}

\usepackage{subfigure}
\usepackage{float} 

\usepackage{authblk}
\title{SWITCHING ACTIVITY IN AN ENSEMBLE OF EXCITABLE NEURONS}
\author[1]{A. G. Korotkov}
\author[1]{S. Y. Zagrebin}
\author[1]{G. V. Osipov}
\affil{Department of Control Theory and System Dynamics, Scientific and Educational Mathematical Center “Mathematics of future technologies”}
\date{}

\graphicspath{ {./fig} }

\begin{document}
\maketitle

\begin{abstract}
The paper proposes two dynamical systems based on the generalized Lotka-Volterra model of three excitable elements interacting through excitatory couplings. It is shown that for some values of the coupling parameters in the phase space of systems, there are heteroclinic cycles containing three or six saddle equilibrium states and heteroclinic curves connecting them. Under certain external stimuli that transfer the system from a stable zero equilibrium state to a small neighborhood of the heteroclinic cycle, the phase trajectory will alternately visit the neighborhood of saddle equilibrium states (possibly more than once), after which it will return to its initial state. The described behavior is proposed to be used to simulate switching activity in neural ensembles. Different transients are determined by different external stimuli. The passage of the phase point of the system near the saddle equilibrium states included in the heteroclinic circuit is proposed to be interpreted as activation of the corresponding element.
\end{abstract}

\section{Introduction}
Often, dynamic processes in neural ensembles are sequential switches of activity between individual neurons and/or groups of neurons. This type of activity is associated with many physiological functions of the nervous system. The generation and propagation of excitation sequences between neurons or groups of them play a crucial role in the brain functioning. For example, sequential activity is observed in neural networks of sensory and motor systems of animals \cite{rabinovich2006dynamical}. In addition, the regions of the brain responsible for the reproduction of singing in the brain of birds is the one that generates sequences of bursting activity \cite{hahnloser2002ultra}. A number of examples where excitation sequences play a key role can be continued \cite{galan2004odor, levi2004dual}.

Currently, an urgent task is to study sequential neural activity through the theory of dynamical systems. In \cite{rabinovich2006dynamical} it was hypothesized that sequential activity may be a consequence of the presence of a stable heteroclinic contour in the phase space of a dynamic system modeling the dynamics of a neural network \cite{guckenheimer1988structurally, stone1990random, postlethwaite2010resonance, driesse2009resonance}. According to it, the basis of the transient dynamics is the winnerless competition principle  \cite{seliger2003dynamics, afraimovich2008winnerless, rabinovich2020sequential}. In short, the essence of this principle is the existence of a stable heteroclinic contour between trajectories of saddle type (these can be saddle equilibrium states, saddle limit cycles, etc.)  \cite{afraimovich2004heteroclinic, afraimovich2004origin, rabinovich2014chunking, latorre2019rhythmic}. When the representative point is in a small neighborhood of the saddle trajectory, it is assumed that a certain element of the network (neuron or group of neurons) corresponding to this saddle trajectory is activated. Thus, a stable heteroclinic contour in phase space is a mathematical image of sequential switching of activity in neural motifs. It was shown in \cite{afraimovich2004heteroclinic} that such a contour exists in ensembles with inhibitory couplings.

Traditionally, in dynamical systems, the main attention is paid to asymptotic modes (at $t \rightarrow \infty$), and transition processes are discarded. However, in neural motifs, the transition process is often more important than the attractor \cite{durstewitz2008computational, muezzinoglu2010transients, brinkman2022metastable}. This fact brings some specifics to the study of models of neural ensembles: for example, it may matter whether it is a stable equilibrium state, a node or a focus (this difference is usually neglected when studying only the asymptotic properties of a dynamical system). The presence/absence of saddle limit cycles, etc., may also be important. 

The main attention in the work will be paid to the study of transition processes associated with the presence of heteroclinic contours: closed contours consisting of one-dimensional heteroclinic trajectories connecting saddle special trajectories. In the models under consideration, these are saddle equilibrium states.

Mathematical models of neural ensembles can be divided into two types: microscopic level models (cell level) and large-scale models (neural ensembles level). In the first type models each neuron of an ensemble is modeled by a system of differential or difference equations, and connections are established between neurons (which can also be modeled by a system of differential or difference equations).

The second type includes Wilson-Cowan models and generalized Lotka-Volterra models. These models describe the interaction between populations of neurons. In this paper, a generalized Lotka-Volterra model is used to model a neural motif.


\section{The first type system}

In this section we will study the dynamics of the activity of neural ensembles consisting of elements modeled by the following equation
\begin{equation} \label{LV_equation1}
\dot \rho = \rho(\rho - 1).
\end{equation}
Here and further, the variable $\rho$ describes the activity of a neuron. 
The neuron is inactive when $\rho=0$  and active when $\rho >0$. The maximum activity takes place at $\rho=1$. The paper assumes to consider the interval of change of $\rho$ from 0 to 1.
The dynamics of the system described by the Eq.~\eqref{LV_equation1} is trivial. The equilibrium state $\rho_1 = 0$ is stable. The equilibrium state $\rho_2 = 1$ is unstable.
At $\rho = 0$, the individual neuron is in an unexcited (inactive) state. In the study of the dynamics of neural ensembles, it will be assumed that all the neurons of the ensemble are not excited at the initial moment of time. The excitation of neurons will occur due to the supply of external stimuli.

We will analyze the dynamics of the ensemble of their three elements. The dynamical system in this case is a special case of the generalized Lotka-Volterra model and has the form

\begin{equation} \label{LV_system1}
\begin{cases}
\dot \rho_1 = \rho_1(\rho_1 - 1 + \alpha \rho_2 + \beta \rho_3) + a\, \delta(\rho_1)\\
\dot \rho_2 = \rho_2(\rho_2 - 1 + \alpha \rho_3 + \beta \rho_1) + b\, \delta(\rho_2)\\
\dot \rho_3 = \rho_3(\rho_3 - 1 + \alpha \rho_1 + \beta \rho_2) + c\, \delta(\rho_3)
\end{cases},
\end{equation}
where $\delta(x)$ is Dirac delta function. $a,b,c,$ parameters are amplitudes of the external stimulus on each element. $\alpha$ and $\beta$ parameters define the couplings between the elements. The paper will consider only the case of $\alpha > 0$, $\beta > 0$, which corresponds to the case of excitatory couplings.

\subsection{A study of a two-dimensional system on an invariant plane $\rho_3 = 0$}
Since one of the conditions $\rho_i = 0$ is met, the $i$-th equation of Eqs.~\eqref{LV_system1} turns into an identity, then the three planes $\rho_i = 0$ are invariant.
To understand the dynamics in a three-dimensional space, we first consider the behavior of two-dimensional systems on the invariant planes $\rho_i = 0$.

On the invariant plane $\rho_3 = 0$ Eqs.~\eqref{LV_system1} takes the form
\begin{equation} \label{LV_system1_2d}
\begin{cases}
\dot \rho_1 = \rho_1(\rho_1 - 1 + \alpha \rho_2)\\
\dot \rho_2 = \rho_2(\rho_2 - 1 + \beta \rho_1)
\end{cases}.
\end{equation}
This system describes the interaction of two interconnected elements.

The equilibrium states of Eqs.~\eqref{LV_system1_2d} and their eigenvalues are given in Table~\ref{table_1}.
\begin{table}[H]
    \centering
    \begin{tabular}{| c | c | c |}
        \hline
        Equilibrium state & \multicolumn{2}{c|}{Eigenvalues}\\
        \hline
        $O_1(0, 0)$ & $-1$ & $-1$\\
        \hline
        $O_2(1, 0)$ & $1$ & $\beta - 1$\\
        \hline
        $O_3(0, 1)$ & $1$ & $\alpha - 1$\\
        \hline
        $O_4(\frac{\alpha - 1}{\alpha \beta - 1}, \frac{\beta - 1}{\alpha \beta - 1})$ & $1$ & $\frac{(1 - \alpha)(\beta - 1)}{\alpha \beta - 1}$\\
        \hline
    \end{tabular}
    \caption{Steady states of Eqs.~\eqref{LV_system1_2d} and it's eigenvalues.}
    \label{table_1}
\end{table}

Let us define the types of equilibrium states of Eqs.~\eqref{LV_system1_2d} depending on the values of the $\alpha$ and $\beta$ parameters.
\begin{enumerate}\itemsep=-2pt
    \item $O_1$ is always a stable node; 
    \item In the region A ($0 < \alpha, \beta < 1$), the equilibria $O_2$ and $O_3$ are saddles, $O_4$ is an unstable node.
    \item In the region B ($0 < \beta < 1 < \alpha$), $O_2$ is a saddle, $O_3$ is an unstable node, $O_4$ has one negative coordinate (at $\alpha \beta \neq 1$).
    \item In the region C ($0 < \alpha < 1 < \beta$), $O_2$ is an unstable node, $O_3$ is a saddle, $O_4$ has one negative coordinate (at $\alpha \beta \neq 1$).
    \item In the region D ($\alpha, \beta > 1$), the equilibria $O_2$ and $O_3$ are unstable nodes, $O_4$ is a saddle.
\end{enumerate}

Since the negative values of $\rho_i$ are not considered, there are three equilibrium states in the regions B and C, and they lie on the $\rho_1$ and $\rho_2$ axes.

\begin{figure}[H]
    \centering
    \includegraphics[width = .5\linewidth]{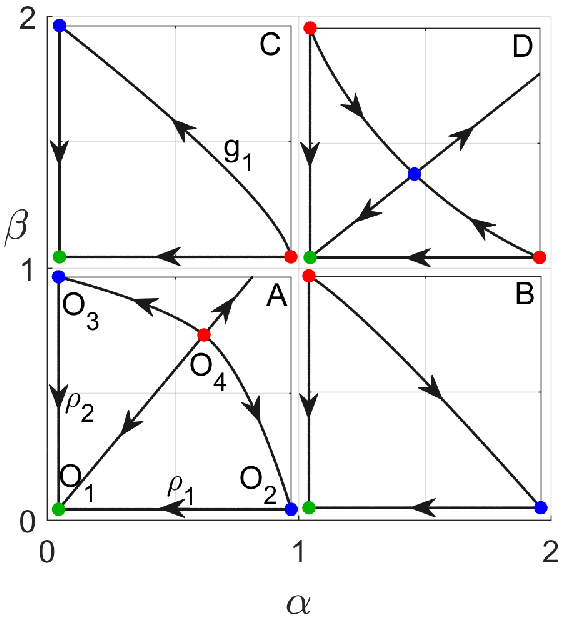}
    \caption{Different types of phase portraits of Eqs.~\eqref{LV_system1_2d}.} In all subsequent figures the following notation is used: blue dots correspond to saddle equilibrium states, green - to stable, red - to unstable.
    \label{PP_bif_diag_syst1_2d}
\end{figure}

Figure~\ref{PP_bif_diag_syst1_2d} shows the partitioning of the parameter plane into the regions A, B, C, D and the corresponding phase portraits. 

Using the Bendixon-Dulac theorem, we show that there are no limit cycles in the phase space of the model \eqref{LV_system1_2d} in the region $\rho_{1, 2}> 0$. Indeed, $$\frac{\partial}{\partial \rho_1} \frac{\dot \rho_1}{\rho_1 \rho_2} + \frac{\partial}{\partial \rho_2} \frac{\dot \rho_2}{\rho_1 \rho_2} = \frac{\partial}{\partial \rho_1} \frac{\rho_1 - 1 + \alpha \rho_2}{\rho_2} + \frac{\partial}{\partial \rho_2} \frac{\rho_2 - 1 + \beta \rho_1}{\rho_1} = \frac{1}{\rho_2} + \frac{1}{\rho_1}.$$ Thus, $\frac{\partial}{\partial \rho_1} \frac{\dot \rho_1}{\rho_1 \rho_2} + \frac{\partial}{\partial \rho_2} \frac{\dot \rho_2}{\rho_1 \rho_2} > 0$ in the region $\rho_{1, 2}> 0$, this means that there are no limit cycles in this region.

Let us consider the parameter region $C$ (similarly, the region $B$ is considered). On the plane ($\rho_1, \rho_2$) there is a trajectory $g_1$ going from the unstable node $O_2$ to the saddle $O_3$. For the three-dimensional system considered below, this trajectory $g_1$ is heteroclinic, since in a three-dimensional system the equilibrium state $O_3$ is a saddle. Similarly, it is possible to show the presence of heteroclinic trajectories on the invariant planes $\rho_1=0$ and $\rho_2=0$. The heteroclinic cycle in a three-dimensional system consists precisely of such heteroclinic trajectories.

The system described by Eqs.~\eqref{LV_system1_2d} has one drawback. For the initial conditions lying below the curve $g_1$, all trajectories come to a stable node $O_1$. And for the initial conditions lying above the curve $g_1$, all trajectories go to infinity. This disadvantage disappears in the second model, discussed below.

\subsubsection{Bifurcations in a two-dimensional system} 
The analysis of the eigenvalues of the equilibria given in Table~\ref{table_1} allows us to determine the bifurcations occurring in Eqs.~\eqref{LV_system1_2d}. At $\beta = 1$, the equilibrium states $O_2$ and $O_4$ undergo transcritical bifurcation. At $\alpha = 1$, the equilibrium states $O_3$ and $O_4$ also undergo transcritical bifurcation. At $\alpha\beta = 1$, the equilibrium state $O_4$ does not exist; when passing through this curve, the sign of one of the coordinates changes to the opposite. The bifurcation diagram of Eqs.~\eqref{LV_system1_2d} is shown in Fig.~\ref{PP_bif_diag_syst1_2d}.

\subsection{Dynamics of a three-neuron ensemble (Eqs.~\eqref{LV_system1})}
Eqs.~\eqref{LV_system1} can have 8 equilibrium states. The coordinates of the equilibrium states and their eigenvalues are given in Table~\ref{table_2}.
\begin{table}[H]
    \centering
    
    \begin{tabular}{| c | c | c | c |}
        \hline
        Equilibrium state & \multicolumn{3}{c|}{Eigenvalues}\\
        \hline
        $O_1(0, 0, 0)$ & $-1$ & $-1$ & $-1$\\
        \hline
        $O_2(1, 0, 0)$ & $1$ & $\alpha - 1$ & $\beta - 1$\\
        \hline
        $O_3(0, 1, 0)$ & $1$ & $\alpha - 1$ & $\beta - 1$\\
        \hline
        $O_4(0, 0, 1)$ & $1$ & $\alpha - 1$ & $\beta - 1$\\
        \hline
        $O_5(0, \frac{\alpha - 1}{\alpha \beta - 1}, \frac{\beta - 1}{\alpha \beta - 1})$ & $1$ & $\frac{\alpha^2 + \beta^2 - \alpha \beta - \alpha - \beta + 1}{\alpha \beta - 1}$ & $\frac{(\alpha - 1)(1 - \beta)}{\alpha \beta - 1}$\\
        \hline
        $O_6(\frac{\beta - 1}{\alpha \beta - 1}, 0, \frac{\alpha - 1}{\alpha \beta - 1})$ & $1$ & $\frac{\alpha^2 + \beta^2 - \alpha \beta - \alpha - \beta + 1}{\alpha \beta - 1}$ & $\frac{(\alpha - 1)(1 - \beta)}{\alpha \beta - 1}$\\
        \hline
        $O_7(\frac{\alpha - 1}{\alpha \beta - 1}, \frac{\beta - 1}{\alpha \beta - 1}, 0)$ & $1$ & $\frac{\alpha^2 + \beta^2 - \alpha \beta - \alpha - \beta + 1}{\alpha \beta - 1}$ & $\frac{(\alpha - 1)(1 - \beta)}{\alpha \beta - 1}$\\
        \hline
        $O_8(\frac{1}{\alpha + \beta + 1}, \frac{1}{\alpha + \beta + 1}, \frac{1}{\alpha + \beta + 1})$ & $1$ & $\frac{2 - \alpha - \beta + (\alpha - \beta) \sqrt{3} i}{2(\alpha + \beta + 1)}$ & $\frac{2 - \alpha - \beta - (\alpha - \beta) \sqrt{3} i}{2(\alpha + \beta + 1)}$\\
        \hline
    \end{tabular}
   \caption{The equilibrium states of Eqs.~\eqref{LV_system1} and their eigenvalues.}
    \label{table_2}
\end{table}

\begin{figure}[H]
    \centering
    \includegraphics[width = .5\linewidth]{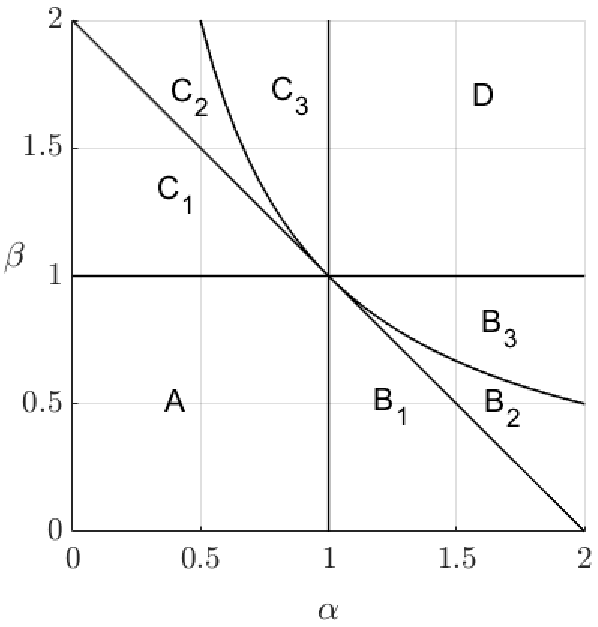}
   \caption{Bifurcation diagram of Eqs.~\eqref{LV_system1}.}
    \label{Bif_diag_1}
\end{figure}

Figure~\ref{Bif_diag_1} shows the bifurcation diagram of Eqs.~\eqref{LV_system1}. 
The following are descriptions of equilibrium states in all regions of the diagram~\ref{Bif_diag_1}.
\begin{enumerate}\itemsep=-2pt
   \item $O_1$ is a stable node.
    \item In the region $A$ ($0 < \alpha, \beta < 1$) the equilibria $O_2$, $O_3$ and $O_4$ are saddles with a two-dimensional stable manifold and a one-dimensional unstable manifold, the equilibria $O_5$, $O_6$ and $O_7$ are saddles with a one-dimensional stable manifold and a two-dimensional unstable manifold, the equilibrium state $O_8$ is an unstable focus (at $2 - \alpha -\beta < 1$) or a node.
    \item In the regions $B_1$ ($\alpha > 1$, $\alpha + \beta < 2$) and $C_1$ ($\beta > 1$, $\alpha +\beta < 2$) the equilibria $O_2$, $O_3$ and $O_4$ are saddles with a two-dimensional unstable manifold and a one-dimensional stable manifold, the equilibria $O_5$, $O_6$ and $O_7$ are saddles with a one-dimensional unstable manifold and a two-dimensional stable manifold, the equilibrium state $O_8$ is an  unstable focus.
   \item In the regions $B_2$ ($\alpha > 1$, $\alpha+\beta > 2$, $\alpha\beta < 1$) and $C_2$ ($\beta > 1$, $\alpha +\beta > 2$, $\alpha\beta < 1$) the equilibria $O_2$, $O_3$ and $O_4$ are saddles with a two-dimensional unstable manifold and a one-dimensional stable manifold, the equilibria $O_5$, $O_6$ and $O_7$ are saddles with a one-dimensional unstable manifold and a two-dimensional stable manifold, the equilibrium state $O_8$ is a saddle-focus.
   \item In the regions $B_3$ ($\beta < 1$, $\alpha\beta > 1$) and $C_3$ ($\alpha <1$, $\alpha\beta > 1$) the equilibria $O_2$, $O_3$ and $O_4$ are saddles with a two-dimensional unstable manifold and a one-dimensional stable manifold, the equilibria $O_5$, $O_6$ and $O_7$ are unstable nodes, the equilibrium state $O_8$ is a saddle-focus.
   \item In the region $D$ ($\alpha, \beta > 1$) the equilibria $O_2$, $O_3$ and $O_4$ are unstable nodes, the equilibria $O_5$, $O_6$ and $O_7$ are saddles with a one-dimensional stable manifold and a two-dimensional unstable manifold, the equilibrium state $O_8$ is a saddle-focus.
\end{enumerate}

When $\alpha\beta = 1$, the equilibria $O_5$, $O_6$, $O_7$ do not exist. In addition, at $\alpha + \beta = 2$ the real part of the two eigenvalues of the equilibrium state $O_8$ turns into $0$. At $\alpha+\beta <2$, this equilibrium state is an unstable saddle-focus with three unstable manifolds, at $\alpha+\beta > 2$, it is a saddle-focus with one-dimensional unstable and two-dimensional stable manifolds. Let's examine this bifurcation in detail.

When the condition is met $\alpha + \beta = 2$ we have $$\dot \rho_1 + \dot \rho_2 + \dot \rho_3 = \rho_1(\rho_1 - 1 + \alpha \rho_2 + (2 - \alpha)\rho_3) + \rho_2(\rho_2 - 1 + \alpha \rho_3 + (2 - \alpha)\rho_1) + \rho_3(\rho_3 - 1 + \alpha \rho_1 + (2 - \alpha)\rho_2) =$$ $$= \rho_1^2 + \rho_2^2 + \rho_3^2 - \rho_1 - \rho_2 - \rho_3 + 2(\rho_1 \rho_3 + \rho_2 \rho_1 + \rho_3 \rho_2) = (\rho_1 + \rho_2 + \rho_3)^2 - \rho_1 - \rho_2 - \rho_3.$$ If $\rho_1 + \rho_2 + \rho_3 = 1$, we get the identity $0 = 0$. Therefore, the plane $\rho_1 + \rho_2 + \rho_3 = 1$ is an invariant set.

On the invariant plane $\rho_1 +\rho_2 + \rho_3 = 1$, provided $\alpha +\beta = 2$, Eqs.~\eqref{LV_system1} takes the form
\begin{equation}
\begin{cases}
\dot \rho_1 = (\alpha - 1)\rho_1(\rho_1 - 1 + 2 \rho_2)\\
\dot \rho_2 = (1 - \alpha)\rho_2(\rho_2 - 1 + 2 \rho_1)
\end{cases}.
\end{equation}
This system can be transformed by substitution of time into
\begin{equation}
\begin{cases}
\dot \rho_1 = -\rho_1(\rho_1 - 1 + 2 \rho_2)\\
\dot \rho_2 = \rho_2(\rho_2 - 1 + 2 \rho_1)
\end{cases}.
\end{equation}
The solution of this system is $$2\rho_1 \rho_2^2 - 2\rho_1 \rho_2(1 - \rho_1) - C = 0.$$
The diagrams of this function inside the region $\rho_{1, 2} > 0$, $\rho_1 + \rho_2 < 1$ are closed curves. Thus, under the condition of $\alpha + \beta = 2$ on the invariant plane $\rho_1 +\rho_2 + \rho_3 = 1$ through each point of the set $\rho_{1, 2, 3} > 0$ a closed curve passes.

\begin{figure}
    \centering
    \subfigure[The region $A$]
    {
        \includegraphics[width = .4\linewidth]{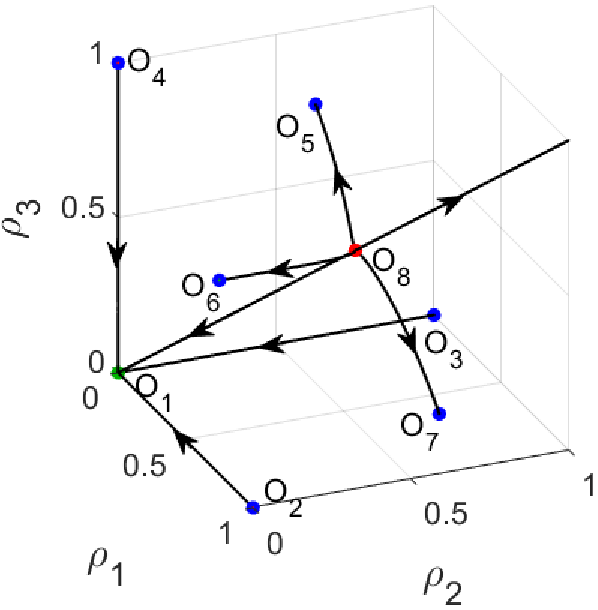}
    }
    \subfigure[The region $C_1$]
    {
        \includegraphics[width = .4\linewidth]{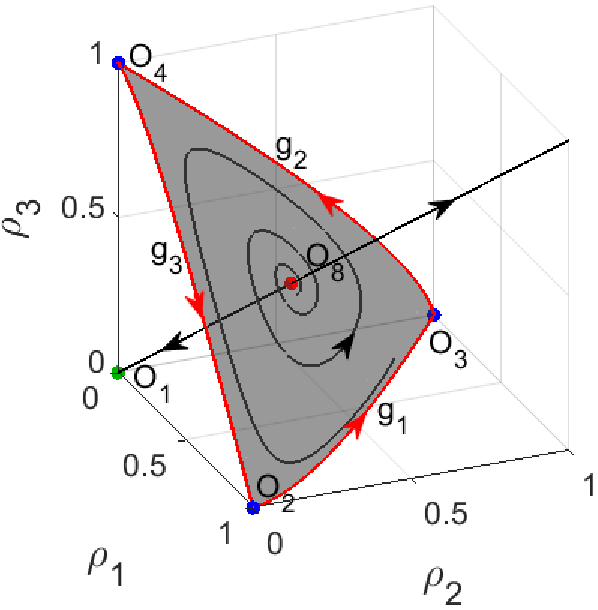}
        \label{PP_syst1_b}
    }\\
    \subfigure[The regions $C_2$ and $C_3$]
    {
        \includegraphics[width = .4\linewidth]{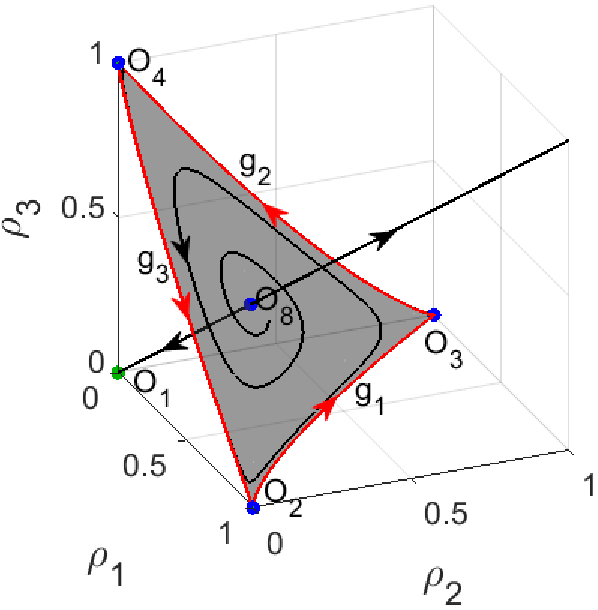}
        \label{PP_syst1_c}
    }
    \subfigure[The region $D$]
    {
        \includegraphics[width = .4\linewidth]{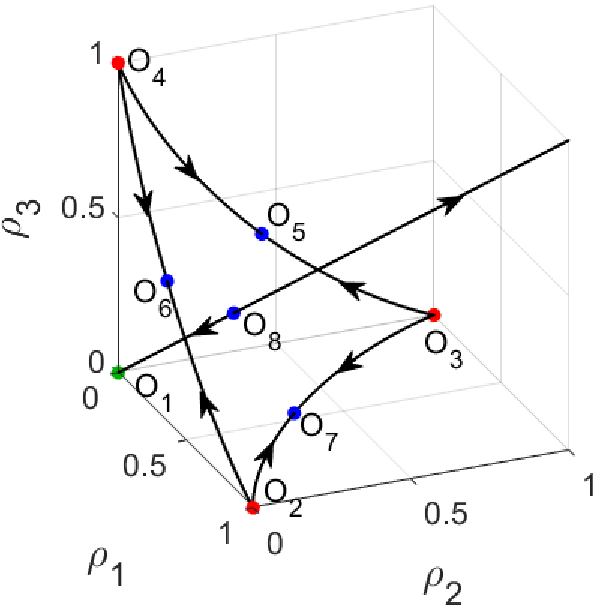}
    }
  \caption{Various types of phase portraits of Eqs.~\eqref{LV_system1}. The heteroclinic trajectories forming the heteroclinic cycle $G$ are shown in red in Figs. b) and c) the invariant manifold $M$ is highlighted in gray. In the region $C_1$, the manifold $M$ is convex, and in the regions $C_2$ and $C_3$ is concave. }
    \label{PP_syst1}
\end{figure}

Figure~\ref{PP_syst1} shows all the different types of phase portraits of Eqs.~\eqref{LV_system1} for parameters from the regions shown in Fig.~\ref{Bif_diag_1}.

The dynamics of an ensemble of three neurons, depending on the parameters, can be as follows:
\begin{enumerate}\itemsep=-2pt
    \item For all regions (Fig.~\ref{PP_syst1}), depending on the external stimulus, the phase trajectories either come to a stable equilibrium state $O_1$ or go to infinity.
    \item For the regions $C_1$, $C_2$ and $C_3$, the boundary between the basins of attraction of the equilibrium and infinity is the integral manifold $M$ containing the saddle points $O_2$, $O_3$ and $O_4$ and the saddle-focus $O_8$.
\end{enumerate}

In the previous section, it was shown that in the regions $C_1$, $C_2$ and $C_3$ in the non-negative quadrant of the invariant plane $\rho_3 = 0$ there are three equilibrium states: a stable node $O_1(0, 0)$, an unstable node $O_3(0,1)$ and a saddle $O_2(1,0)$. When adding the third coordinate $\rho_3$, the equilibrium state $O_1(0, 0,0)$ is a stable node, the equilibrium states $O_3(0,1,0)$ and $O_2(1,0,0)$ are saddles with a two-dimensional unstable manifold and a one-dimensional stable manifold.
Between the saddle equilibrium states $O_2(1, 0, 0)$ and $O_3(0, 1, 0)$ there is a heteroclinic trajectory $g_1$. Due to the symmetry of Eqs.~\eqref{LV_system1}, between pairs of saddles $O_3(0, 1, 0)$, $ O_4(0, 0, 1)$ and $O_4(0, 0, 1)$, $ O_2(1, 0, 0)$ there are also heteroclinic trajectories $g_2$ and $g_3$, respectively. Thus, in the phase space of Eqs.~\eqref{LV_system1} there is a heteroclinic cycle $G$ connecting the saddles $O_2$, $O_3$ and $O_4$. Like the usual limit cycle, the heteroclinic cycle can be stable, unstable and saddle. This is determined by the stability (instability) of the manifolds forming this cycle at the intersection. The heteroclinic cycle $G$ unstable at $\alpha + \beta <2$ (manifolds $\rho_1$, $\rho_2$, $\rho_3$ and $M$ are unstable) and saddle at $\alpha + \beta >2$ (the manifold $M$ is stable).

\subsubsection{Sequential activity}
As mentioned above, one of the behaviors of a neural ensemble is the alternate activation of neurons as a result of a response to an external stimulus, followed by the return of all neurons to their original non-excited state. Mathematical images of such behavior can be phase trajectories passing near the heteroclinic cycle $G$. In the model under consideration, this cycle is a closed trajectory consisting of heteroclinic curves connecting the saddle equilibrium states $O_2$, $O_3$, $O_4$, i.e. the saddles lying on the axes. Only in the regions $B_i$ and $C_i$ these saddles have one-dimensional unstable manifolds that connect them.  
Thus, the sequential activity of neurons can be realized only in the case of asymmetric connections. Note that the sequential neural activity was also observed in ensembles with inhibitory asymmetric connections \cite{komarov2009sequentially, komarov2013heteroclinic, rabinovich2010heteroclinic}.

\begin{figure}[H]
    \centering
    \subfigure[]
    {
        \includegraphics[width = .35\linewidth]{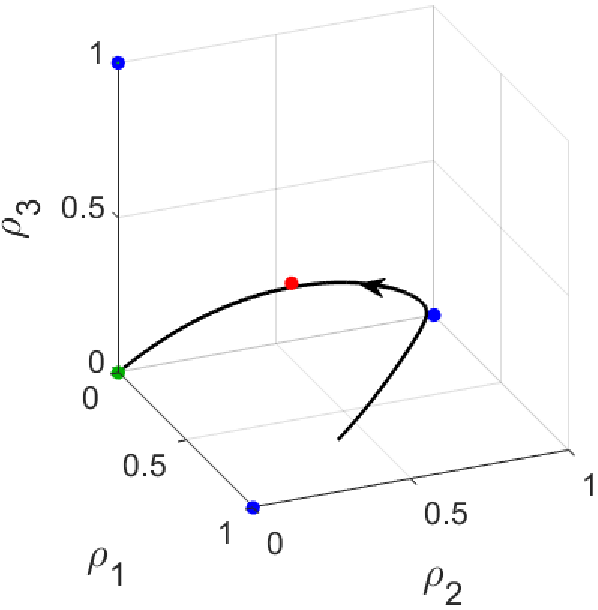}
    }
    \subfigure[]
    {
        \includegraphics[width = .55\linewidth]{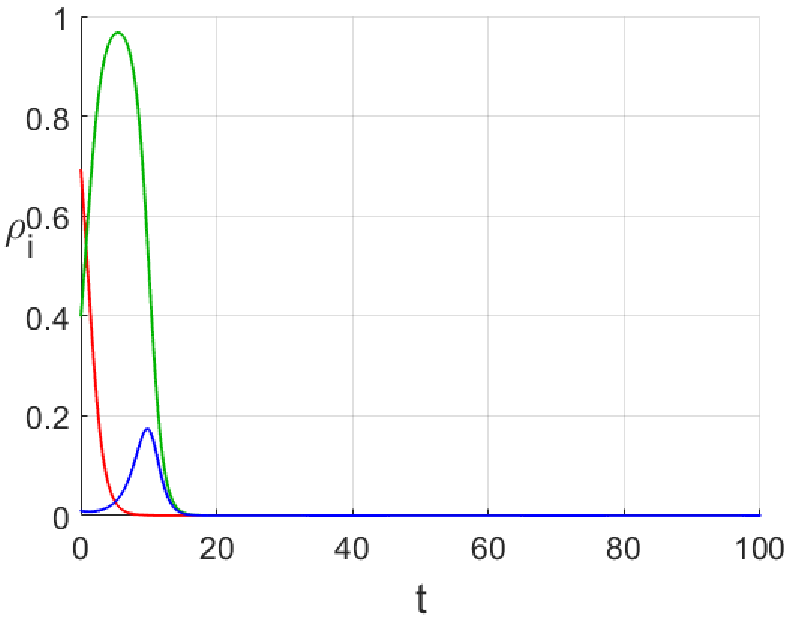}
    }
    \caption{The process of activation and damping. $a$, $b$, $c$ parameters are selected so that only one element is activated - the second one. (a) Phase portrait. (b) Time diagram. Here and further, the red curve corresponds to $\rho_1(t)$, the green curve is $\rho_2(t)$, the blue curve is $\rho_3(t)$. Parameter values: $\alpha = 0.1$, $\beta = 1.5$, $a = 0.6917$, $b = 0.4$, $c = 0.01$.}
    \label{Seq_act_7}
\end{figure}

\begin{figure}[H]
    \centering
    \subfigure[]
    {
        \includegraphics[width = .35\linewidth]{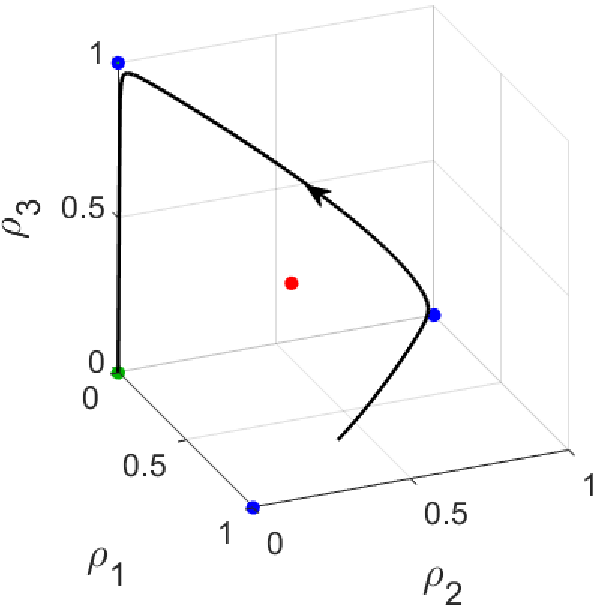}
    }
    \subfigure[]
    {
        \includegraphics[width = .55\linewidth]{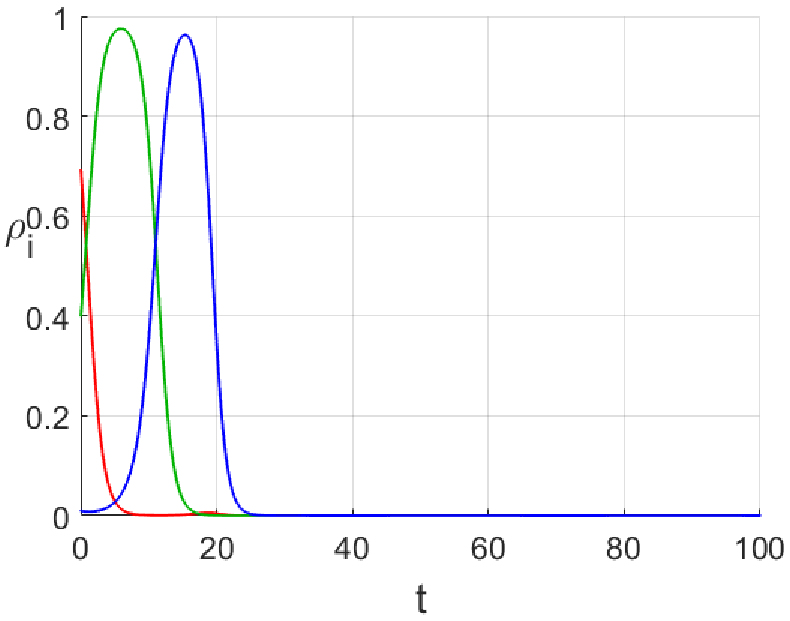}
    }
    \caption{Activation, sequential activity and damping. $a$, $b$, $c$ parameters are selected so that only two elements are activated: the second and the third ones. (a) Phase portrait. (b) Time diagram. Parameter values: $\alpha = 0.1$, $\beta = 1.5$, $a = 0.69173179$, $b = 0.4$, $c = 0.01$.}
\label{Seq_act_6}
\end{figure}

\begin{figure}[H]
    \centering
    \subfigure[]
    {
        \includegraphics[width = .35\linewidth]{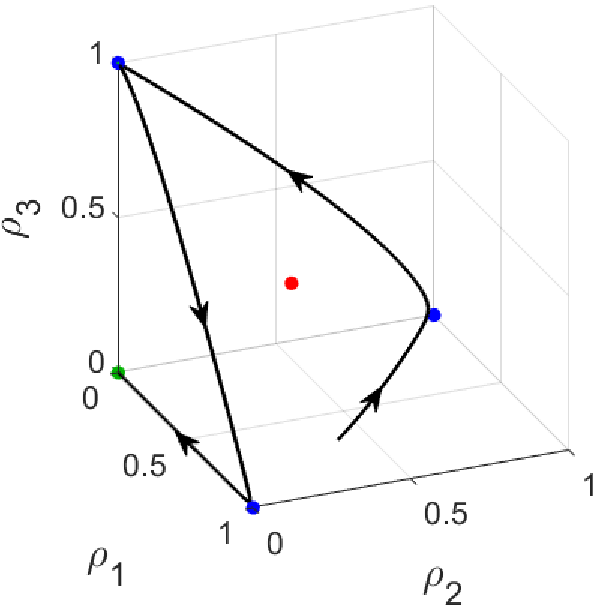}
    }
    \subfigure[]
    {
        \includegraphics[width = .55\linewidth]{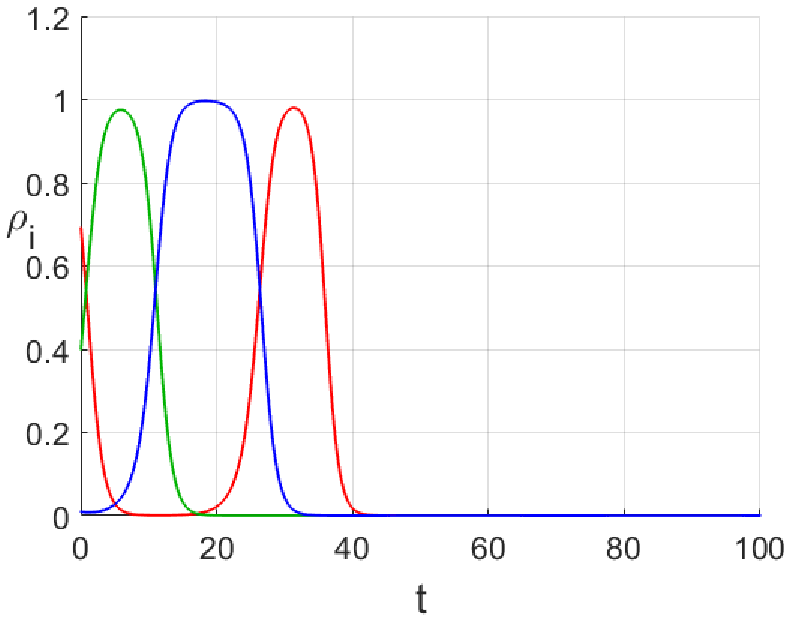}
    }
    \caption{Activation, sequential activity and damping. All three neurons are activated sequentially once. (a) Phase portrait: the phase trajectory takes place in a vicinity of the heteroclinic cycle due to the selection of an external stimuli. (b) Time diagram. Parameter values: $\alpha = 0.1$, $\beta = 1.5$, $a = 0.691731794346462$, $b = 0.4$, $c = 0.01$.}
    \label{Seq_act_1}
\end{figure}

\begin{figure}[H]
    \centering
    \subfigure[]
    {
        \includegraphics[width = .35\linewidth]{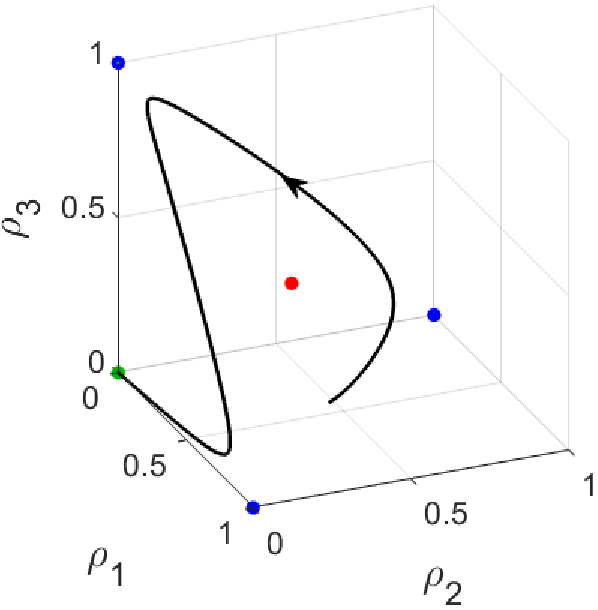}
    }
    \subfigure[]
    {
        \includegraphics[width = .55\linewidth]{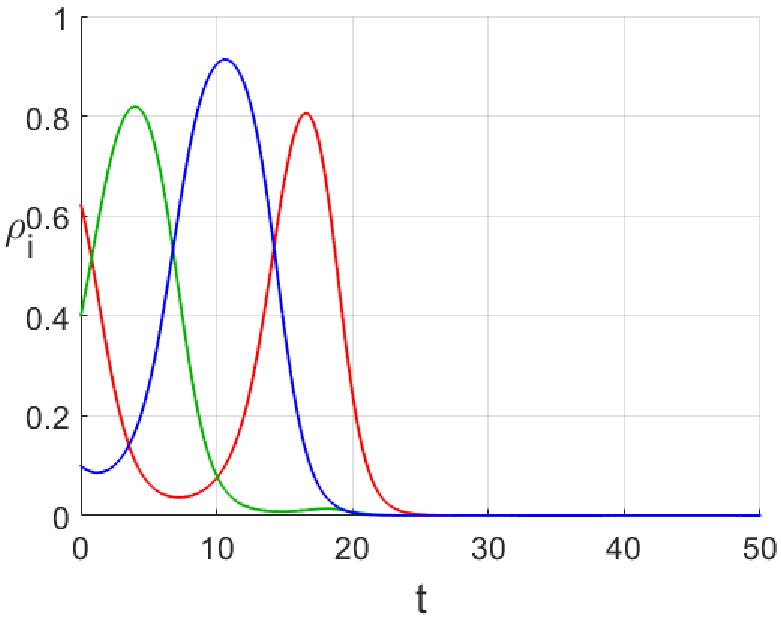}
    }
    \caption{Activation, sequential activity and damping.  All three neurons are activated sequentially once. (a) Phase portrait: the phase trajectory does not take place in a vicinity of the heteroclinic cycle due to the selection of an external stimuli. (b) Time diagram. Parameter values: $\alpha = 0.1$, $\beta = 1.5$,  $\alpha + \beta < 2$, $a = 0.62313692$, $b = 0.4$, $c = 0.1$.}
    \label{Seq_act_3}
\end{figure}

\begin{figure}[H]
    \centering
    \subfigure[]
    {
        \includegraphics[width = .35\linewidth]{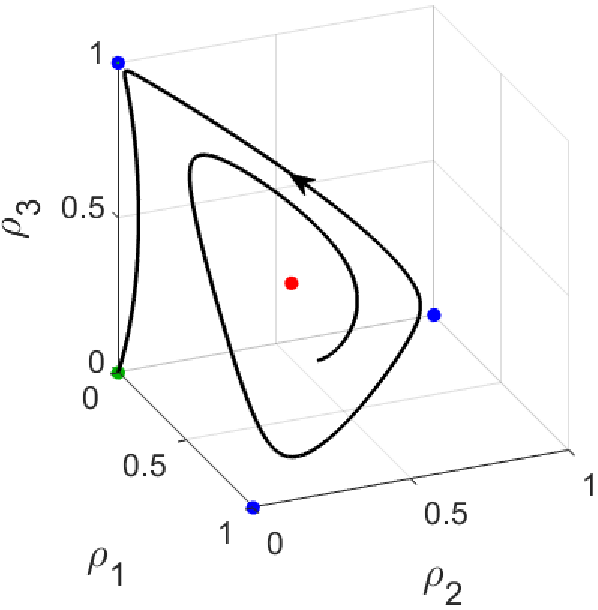}
    }
    \subfigure[]
    {
        \includegraphics[width = .55\linewidth]{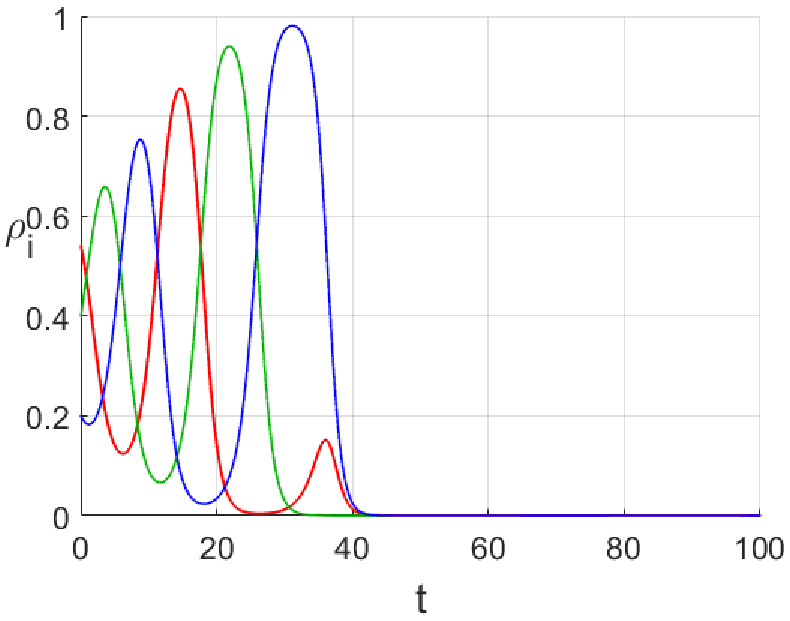}
    }
    \caption{Activation, sequential activity and damping. The second and third neurons are activated twice. The first neuron is activated once. (a) Phase portrait: the phase trajectory takes place in a vicinity of the heteroclinic cycle due to the selection of an external stimuli moving the starting point to the integral manifold $M$. (b) Time diagram. Parameter values: $\alpha = 0.1$, $\beta = 1.5$, $a = 0.540537789125142$, $b = 0.4$, $c = 0.2$.}
    \label{Seq_act_4}
\end{figure}

\begin{figure}[H]
    \centering
    \subfigure[]
    {
        \includegraphics[width = .35\linewidth]{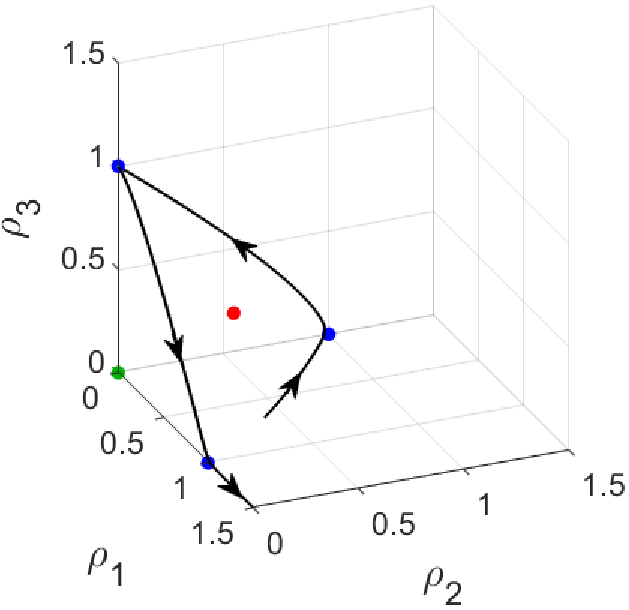}
    }
    \subfigure[]
    {
        \includegraphics[width = .55\linewidth]{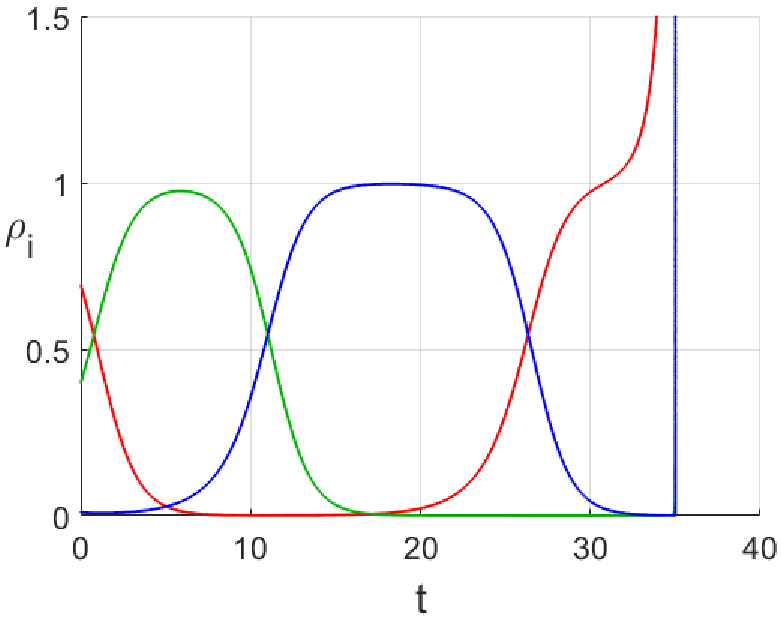}
    }
    \caption{Sequential activity. The phase trajectory goes to infinity. (a) Phase portrait. (b)  Time diagram. Parameter values: $\alpha = 0.1$, $\beta = 1.5$, $a = 0.691731794346463$, $b = 0.4$, $c = 0.01$.}
    \label{Seq_act_2}
\end{figure}

\begin{figure}[H]
    \centering
    \subfigure[]
    {
        \includegraphics[width = .35\linewidth]{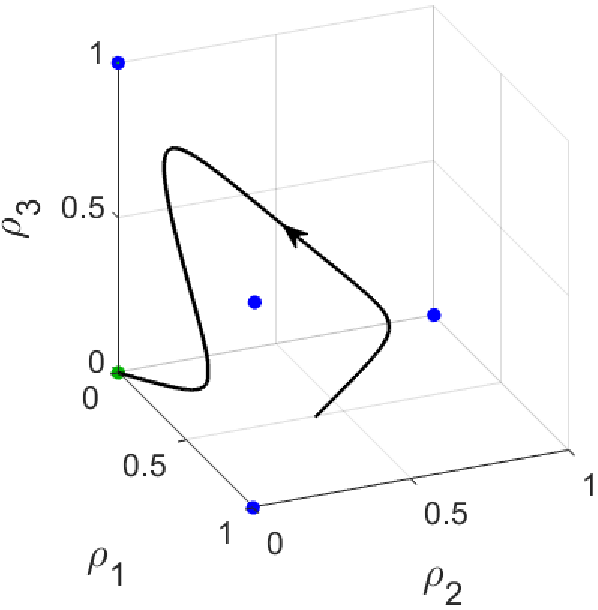}
    }
    \subfigure[]
    {
        \includegraphics[width = .55\linewidth]{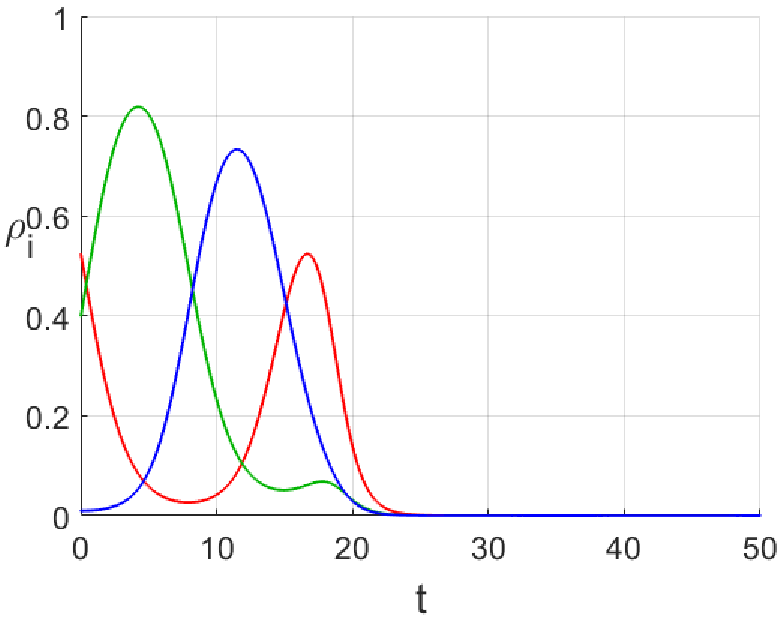}
    }
    \caption{Activation, sequential activity and damping.(a) Phase portrait. (b)  Time diagram. Parameter values: $\alpha = 0.4$, $\beta = 1.9$, $\alpha + \beta >  2$, $a = 0.52382217$, $b = 0.4$, $c = 0.01$.}
    \label{Seq_act_5}
\end{figure}

Further, Figs.~\ref{Seq_act_7}-\ref{Seq_act_5} show phase portraits and alternating realizations of several possible behaviors of a neural ensemble depending on the external stimulus determined by the values $a$, $b$ and $c$.
We will assume that the $i$-th neuron is active if the value of $\rho_i$ exceeds the value of 0.5. Initially, all neurons are inactive. The image point is in the equilibrium state $O_1$. In order to activate the neuron, you need to send such an external stimulus to transfer the initial phase point from the equilibrium state $O_1$ to a small neighborhood of an unstable or saddle heteroclinic cycle $G$ or to a small neighborhood of an invariant manifold $M$ in the case of a saddle heteroclinic cycle $G$.
Depending on the external stimulus, the following dynamics can be obtained:

\begin{itemize}\itemsep=-2pt
    \item excitation of a single neuron (Fig.~\ref{Seq_act_7})
    \item sequential excitation of two neurons (Fig.~\ref{Seq_act_6})
    \item sequential one-time excitation of all three neurons (Figs.~\ref{Seq_act_1},  \ref{Seq_act_3} and \ref{Seq_act_5})
    \item sequential multiple excitation of all three neurons (Fig.~\ref{Seq_act_4})
    \item phase trajectory goes to infinity (Fig.~\ref{Seq_act_2}) 
\end{itemize}

\section{The second type system}

To eliminate the disadvantage of the first system, the possibility of trajectories going to infinity, in this section we will analyze the dynamics of the ensemble modeled by the following equation
\begin{equation} \label{LV_equation2}
\dot \rho = \rho \left(\gamma - \rho \right)(\rho - 1).
\end{equation}
The $\gamma$ parameter is the threshold value of the ensemble's activity. It's value satisfies the inequalities $0 < \gamma < 1$.

Eqs.~\eqref{LV_equation2} has three equilibrium states: $\rho = 0$ and $\rho = 1$ are stable and $\rho =\gamma$ is unstable. If the initial condition is $\rho_0 < \gamma$, then the activity of the $\rho$ ensemble decreases and tends to zero. If $\rho_0 > \gamma$, then the activity of the $\rho$ ensemble increases and tends to one. It follows from the above that the set $0\leq\rho\leq 1$ is invariant.

The model of an ensemble of three connected elements modeled by Eq.~\eqref{LV_equation2} has the form
\begin{equation} \label{LV_system2}
\begin{cases}
\dot{\rho_1} = \rho_1(\gamma - \rho_1)(\rho_1 - 1 + \alpha \rho_2 + \beta \rho_3) + a\,\delta(\rho_1)\\
\dot{\rho_2} = \rho_2(\gamma - \rho_2)(\rho_2 - 1 + \alpha \rho_3 + \beta \rho_1) + b\,\delta(\rho_2)\\
\dot{\rho_3} = \rho_3(\gamma - \rho_3)(\rho_3 - 1 + \alpha \rho_1 + \beta \rho_2) + c\,\delta(\rho_3)
\end{cases}.
\end{equation}

\subsection{Study of a two-dimensional system on an invariant plane $\rho_3 = 0$}
Since when one of the conditions $\rho_i = 0$ is met, the i-th equation of Eqs.~\eqref{LV_system2} turns into an identity, then the three planes $\rho_i = 0$ are invariant.

On the invariant plane $\rho_3 = 0$, Eqs.~\eqref{LV_system2} takes the form
\begin{equation} \label{LV_system2_2d}
\begin{cases}
\dot \rho_1 = \rho_1(\gamma - \rho_1)(\rho_1 - 1 + \alpha \rho_2) + a\,\delta(\rho_1)\\
\dot \rho_2 = \rho_2(\gamma - \rho_2)(\rho_2 - 1 + \beta \rho_1) + b\,\delta(\rho_2)
\end{cases}.
\end{equation}

The equilibrium states of Eqs.~\eqref{LV_system2_2d} and their eigenvalues are given in Table~\ref{table_3}:
\renewcommand{\arraystretch}{1}
\begin{table}
    \centering
    \begin{tabular}{| c | c | c |}
        \hline
        Equilibrium state & \multicolumn{2}{c|}{Eigenvalues}\\
        \hline
        $O_1(0, 0)$ & $-\gamma$ & $-\gamma$\\
        \hline
        $O_2(1, 0)$ & $\gamma(\beta - 1)$ & $\gamma - 1$\\
        \hline
        $O_3(0, 1)$ & $\gamma(\alpha - 1)$ & $\gamma - 1$\\
        \hline
        $O_4(1 - \alpha \gamma, \gamma)$ & $\gamma(\beta(\alpha \gamma - 1) - \gamma + 1)$ & $(1 - \alpha \gamma)(\gamma + \alpha \gamma - 1)$\\
        \hline
        $O_5(\gamma, 1 - \beta \gamma)$ & $\gamma(\alpha(\beta \gamma - 1) - \gamma + 1)$ & $(1 - \beta \gamma)(\gamma + \beta \gamma - 1)$\\
        \hline
        $O_6(\gamma, 0)$ & $\gamma(\beta \gamma - 1)$ & $\gamma(1 - \gamma)$\\
        \hline
        $O_7(0, \gamma)$ & $\gamma(\alpha \gamma - 1)$ & $\gamma(1 - \gamma)$\\
        \hline
        $O_8(\gamma, \gamma)$ & $\gamma(1 - \gamma - \alpha \gamma)$ & $\gamma(1 - \gamma - \beta \gamma)$\\
        \hline
        $O_9(\frac{\alpha - 1}{\alpha \beta - 1}, \frac{\beta - 1}{\alpha \beta - 1})$ & $\lambda_1$ & $\lambda_2$\\
        \hline
    \end{tabular}
     \caption{The equilibrium states of Eqs.~\eqref{LV_system2_2d} and their eigenvalues.}
     \label{table_3}
\end{table}
\renewcommand{\arraystretch}{1}

The eigenvalues $\lambda_{1, 2}$ of the equilibrium state $O_9$ are too cumbersome and therefore are not given in the table. They are found from the characteristic equation $${\textstyle \lambda^2 + \frac{(\alpha - 1)(\alpha + \gamma - \alpha \beta \gamma - 1) + (\beta - 1)(\beta + \gamma - \alpha \beta \gamma - 1)}{(\alpha \beta - 1)^2}\lambda - \frac{(\alpha - 1)(\alpha + \gamma - \alpha \beta \gamma - 1)(\beta - 1)(\beta + \gamma - \alpha \beta \gamma - 1)}{(\alpha \beta - 1)^3} = 0.}$$

$O_9$ undergoes Andronov-Hopf bifurcation under the following conditions.
\begin{equation}
\label{bifurcation}
\begin{cases}
\alpha \beta \gamma(\alpha + \beta - 2) - \alpha^2 - \beta^2 + (2 - \gamma)(\alpha + \beta) + 2\gamma - 2  = 0\\
\frac{\left(\alpha - 1 \right) \left(\beta - 1 \right) \left(\alpha \beta \gamma(\alpha \beta \gamma - \alpha - \beta - 2\gamma + 2) + (\gamma - 1)(\alpha + \beta) + \alpha \beta + \gamma^2 - 2\gamma + 1 \right)}{(1 - \alpha \beta)^3} > 0
\end{cases}.
\end{equation}
$\beta$ parameter can be expressed from the first equation:
$$\beta = \frac{-\alpha^2 \gamma + 2\alpha \gamma + \gamma - 2 \pm (\alpha - 1)\sqrt{\alpha^2 \gamma^2 + 2\alpha \gamma(2 - \gamma) + \gamma^2 + 4\gamma - 4}}{2(\alpha \gamma - 1)}.$$

\begin{figure}[H]
    \centering
    \subfigure[$\beta = 1.3055$]
    {
        \includegraphics[width = .3\linewidth]{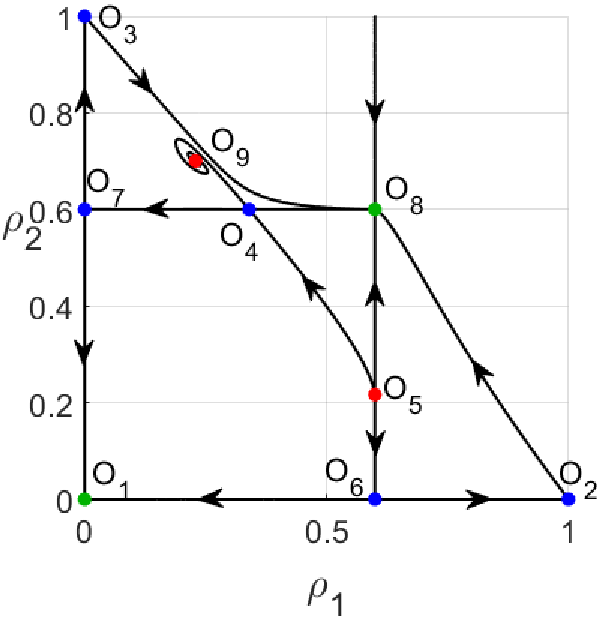}
    }
    \subfigure[$\beta = 1.324802754125605$]
    {
        \includegraphics[width = .3\linewidth]{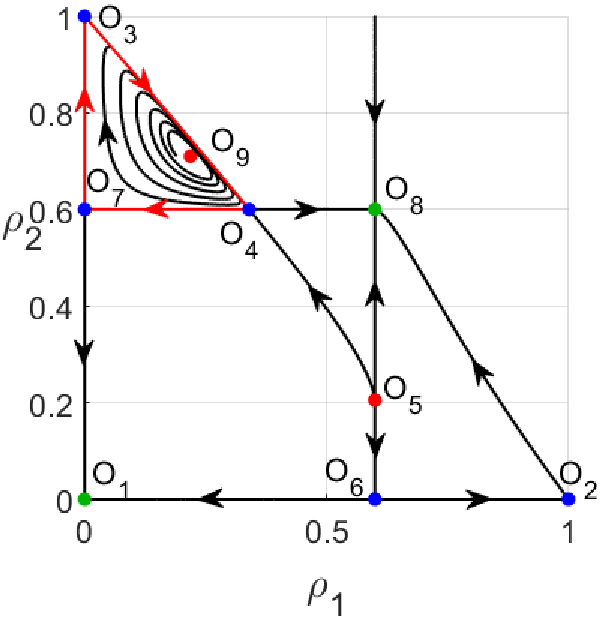}
    }\\
    \subfigure[$\beta = 1.3325$]
    {
        \includegraphics[width = .3\linewidth]{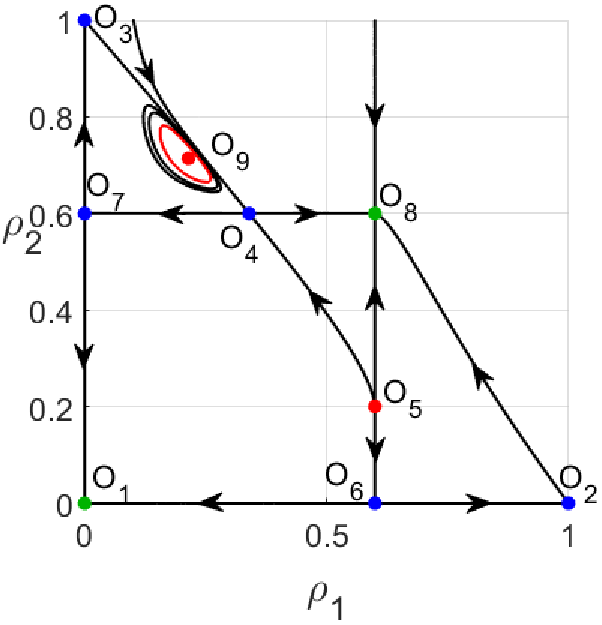}
    }
    \subfigure[$\beta = 1.34$]
    {
        \includegraphics[width = .3\linewidth]{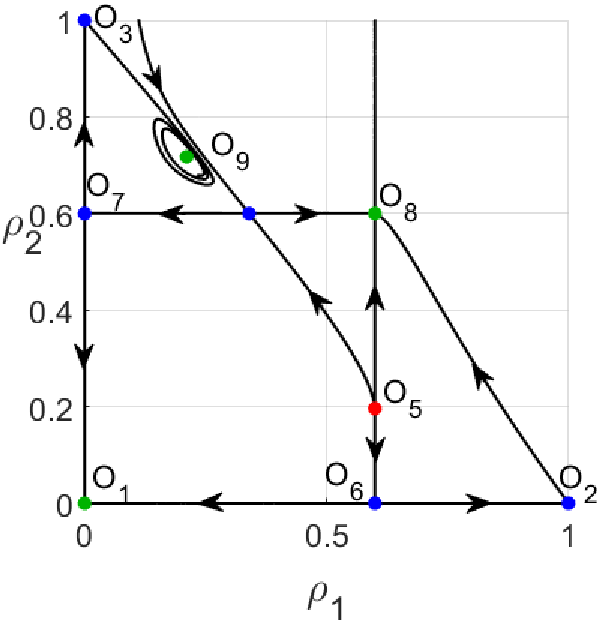}
    }
    \caption{Andronov–Hopf bifurcation and heteroclinic bifurcation. In Figure (b), the heteroclinic cycle connecting the saddles $O_3$, $O_4$ and $O_7$ is highlighted in red. In Figure (c), a stable limit cycle is highlighted in red. In Figure (d), the limit cycle does not exist, while $O_9$ is stable. Parameter values: $\alpha = 1.1$, $\gamma = 0.6$.}
    \label{bif_AH}
\end{figure}

At the intersection of this curve in the direction of decreasing $\beta$ parameter, a stable limit cycle is born in the phase space, which further increases in size and gets into the heteroclinic contour formed by the saddles  $O_3$, $O_4$ and $O_7$ (Fig.~\ref{bif_AH}).

Similarly with the study of the first type system on invariant planes $\rho_i=0$, we determine the types of equilibrium states of Eqs.~\eqref{LV_system2_2d} depending on the values of the $\alpha$ and $\beta$ parameters. 
The number of regions of the parameter space with different number and type of equilibrium states in this model is greater. It is enough to describe only those regions of the parameter space that lie either above the diagonal $\alpha = \beta$, or intersect it, since in the regions symmetrical with respect to the diagonal, the phase portraits differ only in the numbering of the variables $\rho_1$ and $\rho_2$.
\begin{enumerate}\itemsep=-2pt
    \item $O_1$ is always a stable node.
    \item $O_2$ is a stable node in the regions $A$, $B$, $C$, $H$, $I$ ($0 < \beta < 1$) and a saddle in the remaining regions.
    \item $O_3$ is a stable node in the regions $A$, $B$, $C$, $D$, $E$, $F$, $G$, $H$, $I$, $J$, $K$, $L$ ($0 < \alpha < 1$) and a saddle in the remaining regions.
    \item $O_4$ is a stable node in the region $A$ and has one negative coordinate in the region $S$, a saddle in the regions $B$, $C$, $D$, $E$, $F$, $G$, $Q$, $R$, and an unstable node in the remaining regions.
    \item $O_5$ is stable node in the regions $A$, $B$ and has one negative coordinate in the regions $F$, $G$, $L$, $N$, $P$, $R$, $S$, an unstable node in the region $H$ and a saddle in the remaining regions.
    \item $O_6$ is an unstable node in the regions $F$, $G$, $L$, $N$, $P$, $R$, $S$ ($\beta > \frac{1}{\gamma}$) and a saddle in the remaining regions.
    \item $O_7$ is an unstable node in the region $S$ ($\alpha > \frac{1}{\gamma}$) and a saddle in the remaining regions.
    \item $O_8$ is an unstable node in the regions $A$ and $B$, a saddle in the regions $C$, $D$, $E$, $F$ and $G$ and a stable node in the remaining regions.
    \item $O_9$ is stable node in the region $A$, a saddle in the regions $B$, $C$, $I$, $Q$, $R$ and $S$, an unstable node in the region $H$, a stable focus in the regions $M$ and $N$, an unstable focus in the regions $O$ and $P$ and has one negative coordinate in the remaining regions.
\end{enumerate}

Table~\ref{table_4} provides a summary list of the types of all nine equilibria and the regions in which they exist.
\renewcommand{\arraystretch}{1}
\begin{table}[H]
    \centering
    \begin{tabular}{| c | c | c | c | c | c | c | c | c | c |}
        \hline
        Region & $O_1$ & $O_2$ & $O_3$ & $O_4$ & $O_5$ & $O_6$ & $O_7$ & $O_8$ & $O_9$\\
        \hline
        $A$ & SN & SN & SN & SN & SN & S & S & UN & SN\\
        \hline
        $B$ & SN & SN & SN & S & SN & S & S & UN & S\\
        \hline
        $C$ & SN & SN & SN & S & S & S & S & S & S\\
        \hline
        $D$ & SN & S & SN & S & S & S & S & S & -\\
        \hline
        $E$ & SN & S & SN & S & S & S & S & S & -\\
        \hline
        $F$ & SN & S & SN & S & - & UN & S & S & -\\
        \hline
        $G$ & SN & S & SN & S & - & UN & S & S & -\\
        \hline
        $H$ & SN & SN & SN & UN & UN & S & S & SN & UN\\
        \hline
        $I$ & SN & SN & SN & UN & S & S & S & SN & S\\
        \hline
        $J$ & SN & S & SN & UN & S & S & S & SN & -\\
        \hline
        $K$ & SN & S & SN & UN & S & S & S & SN & -\\
        \hline
        $L$ & SN & S & SN & UN & - & UN & S & SN & -\\
        \hline
        $M$ & SN & S & S & UN & S & S & S & SN & SF\\
        \hline
        $N$ & SN & S & S & UN & - & UN & S & SN & SF\\
        \hline
        $O$ & SN & S & S & UN & S & S & S & SN & UF\\
        \hline
        $P$ & SN & S & S & UN & - & UN & S & SN & UF\\
        \hline
        $Q$ & SN & S & S & S & S & S & S & SN & S\\
        \hline
        $R$ & SN & S & S & S & - & UN & S & SN & S\\
        \hline
        $S$ & SN & S & S & - & - & UN & UN & SN & S\\
        \hline
    \end{tabular}
    \caption{Types of equilibrium states in all regions of the parameter space (the partitioning into regions is shown in Fig.~\ref{Bif_diag_syst2_2d}). A dash stands for a presence of a negative coordinate of the equilibrium. SN - stable node, SF - stable focus, UN - unstable node, UF - unstable focus, S - saddle.}
    \label{table_4}
\end{table}
\renewcommand{\arraystretch}{1}

In the system under consideration, two bifurcations of equilibrium states are possible: the Andronov-Hopf bifurcation described above and the transcritical bifurcation. Table~\ref{table_5} shows a list of transcritical bifurcations of the equilibrium states of the system.

\begin{table}
    \centering
    \begin{tabular}{| c | c | c | c | c | c | c | c | c | c |}
        \hline
         & $O_1$ & $O_2$ & $O_3$ & $O_4$ & $O_5$ & $O_6$ & $O_7$ & $O_8$ & $O_9$\\
        \hline
        $O_1(0, 0)$ & - & - & - & - & - & - & - & - & -\\
        \hline
        $O_2(1, 0)$ & - & - & - & - & - & - & - & - & $\beta = 1$\\
        \hline
        $O_3(0, 1)$ & - & - & - & - & - & - & - & - & $\alpha = 1$\\
        \hline
        $O_4(1 - \alpha \gamma, \gamma)$ & - & - & - & - & - & - & $\alpha = \frac{1}{\gamma}$ & $\alpha = \frac{1}{\gamma} - 1$ & $\beta = \frac{\gamma - 1}{\alpha \gamma - 1}$\\
        \hline
        $O_5(\gamma, 1 - \beta \gamma)$ & - & - & - & - & - & $\beta = \frac{1}{\gamma}$ & - & $\beta = \frac{1}{\gamma} - 1$ & $\beta = \frac{\alpha + \gamma - 1}{\alpha \gamma}$\\
        \hline
        $O_6(\gamma, 0)$ & - & - & - & - & $\beta = \frac{1}{\gamma}$ & - & - & - & -\\
        \hline
        $O_7(0, \gamma)$ & - & - & - & $\alpha = \frac{1}{\gamma}$ & - & - & - & - & -\\
        \hline
        $O_8(\gamma, \gamma)$ & - & - & - & $\alpha = \frac{1}{\gamma} - 1$ & $\beta = \frac{1}{\gamma} - 1$ & - & - & - & -\\
        \hline
        $O_9(\frac{\alpha - 1}{\alpha \beta - 1}, \frac{\beta - 1}{\alpha \beta - 1})$ & - & $\beta = 1$ & $\alpha = 1$ & $\beta = \frac{\gamma - 1}{\alpha \gamma - 1}$ & $\beta = \frac{\alpha + \gamma - 1}{\alpha \gamma}$ & - & - & - & -\\
        \hline
    \end{tabular}
    \caption{Lines on which transcritical bifurcations of equilibrium states occur.} 
    \label{table_5}
\end{table}

\begin{figure}[H]
    \centering
    \includegraphics[width = .5\linewidth]{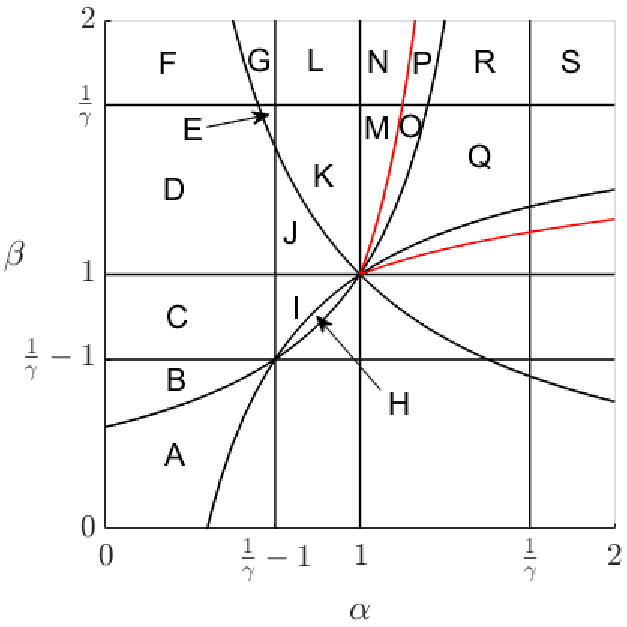}
    \caption{The bifurcation diagram of Eqs.~\eqref{LV_system2_2d} at $\gamma = 0.6$. The black lines correspond to transcritical bifurcations, the red line corresponds to bifurcations of \eqref{bifurcation}. }
    \label{Bif_diag_syst2_2d}
\end{figure}

In Figure \ref{Bif_diag_syst2_2d} the bifurcation diagram of the system described by Eqs.~\eqref{LV_system2_2d} is given for $\gamma=0.6$.

\begin{figure}[H]
    \centering
    \subfigure[Region A]
    {
        \includegraphics[width = .23\linewidth]{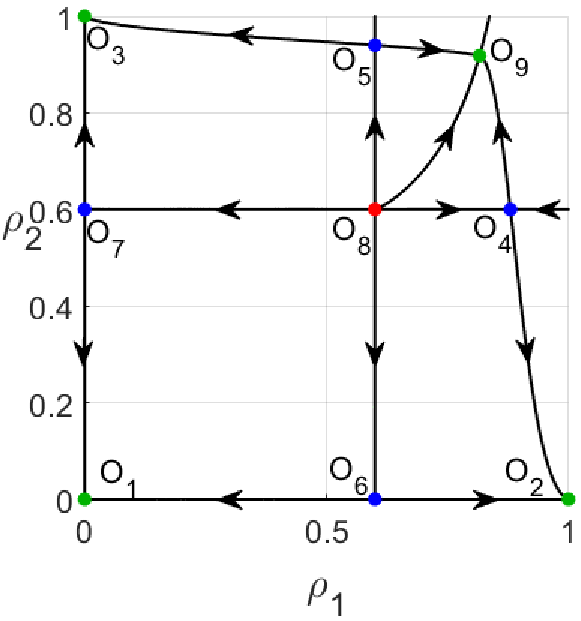}
    }
    \subfigure[Region B]
    {
        \includegraphics[width = .23\linewidth]{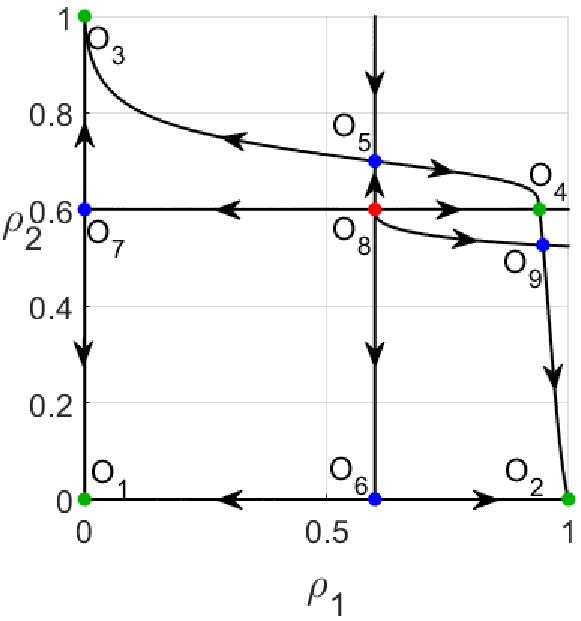}
    }
    \subfigure[Region C]
    {
        \includegraphics[width = .23\linewidth]{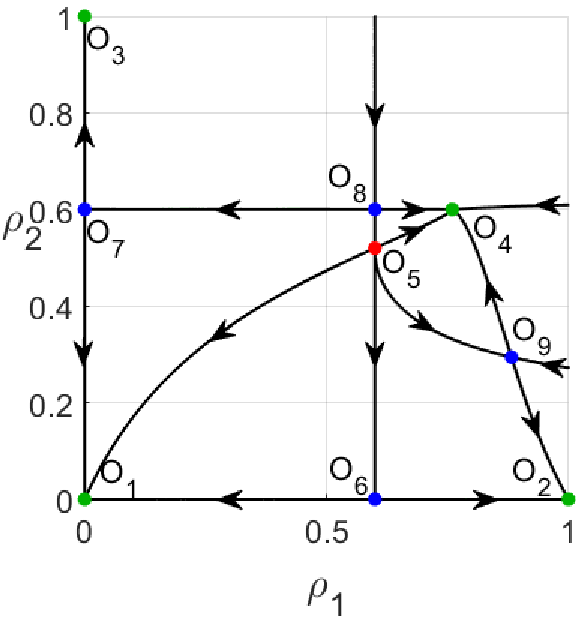}
    }
    \subfigure[Regions D and E]
    {
        \includegraphics[width = .23\linewidth]{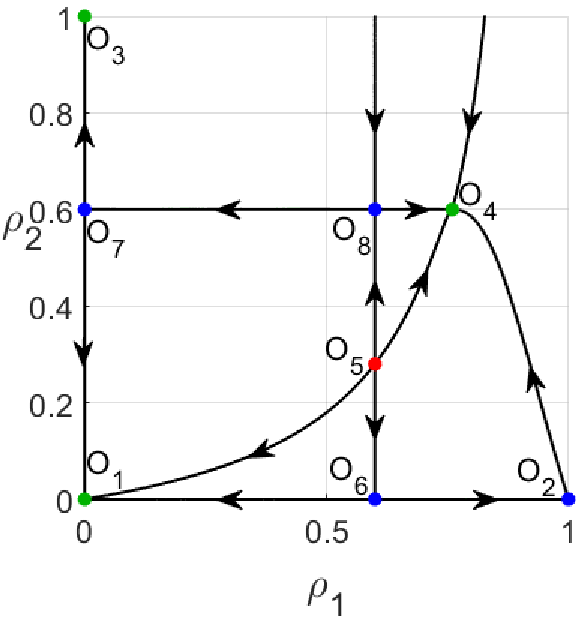}
    }\\
    \subfigure[Regions F and G]
    {
        \includegraphics[width = .23\linewidth]{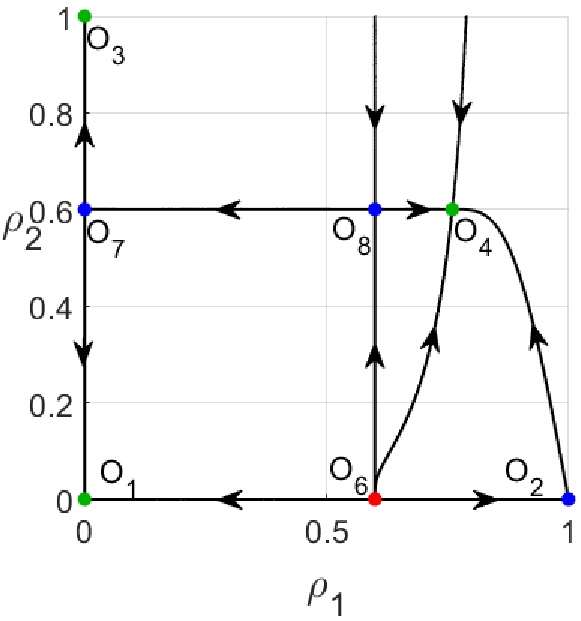}
    }
    \subfigure[Region H]
    {
        \includegraphics[width = .23\linewidth]{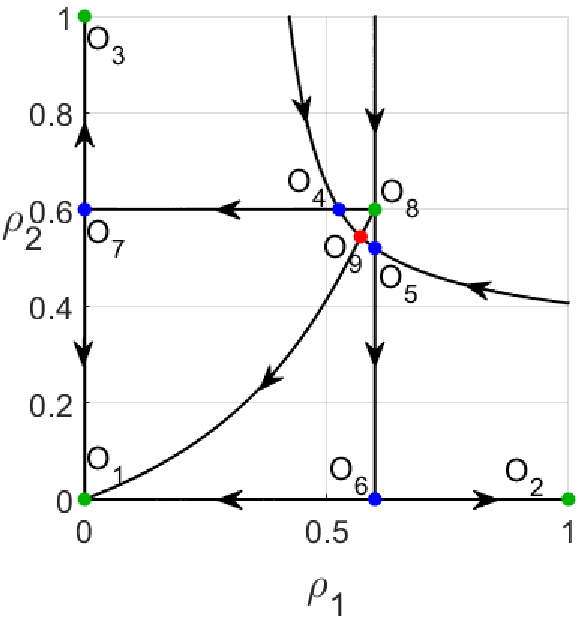}
    }
    \subfigure[Region I]
    {
        \includegraphics[width = .23\linewidth]{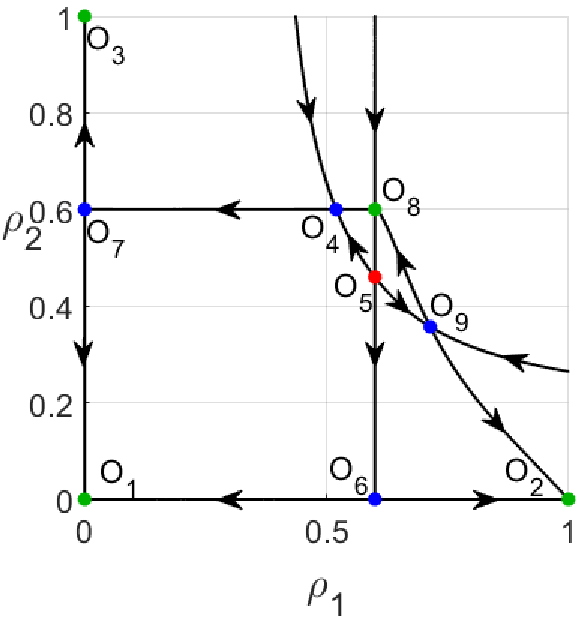}
    }
    \subfigure[Regions J and K]
    {
        \includegraphics[width = .23\linewidth]{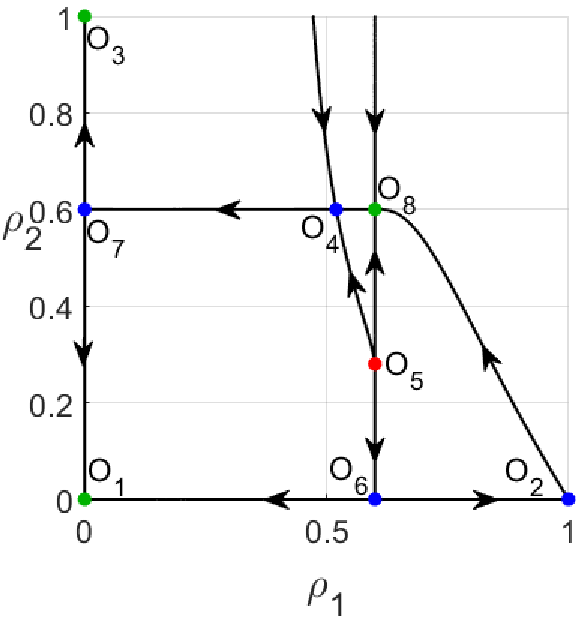}
    }\\
    \subfigure[Region L]
    {
        \includegraphics[width = .23\linewidth]{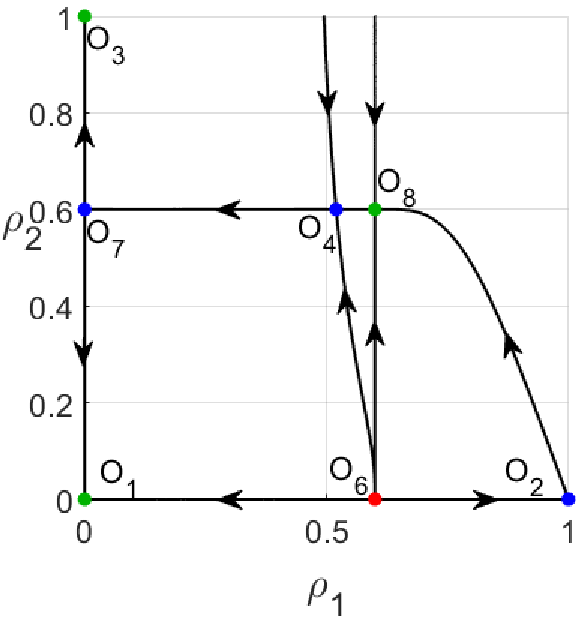}
    }
    \subfigure[Region M]
    {
        \includegraphics[width = .23\linewidth]{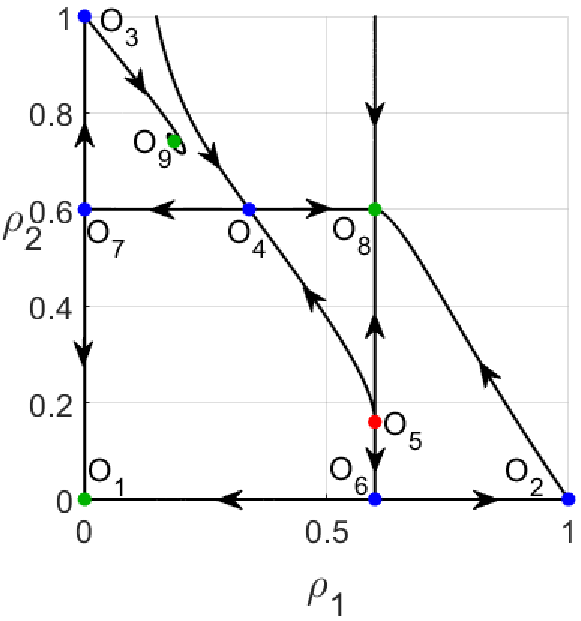}
    }
    \subfigure[Region N]
    {
        \includegraphics[width = .23\linewidth]{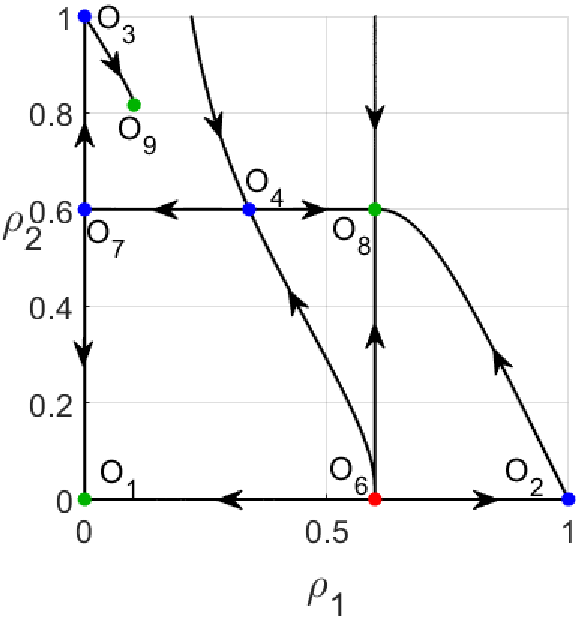}
    }
    \subfigure[Region O]
    {
        \includegraphics[width = .23\linewidth]{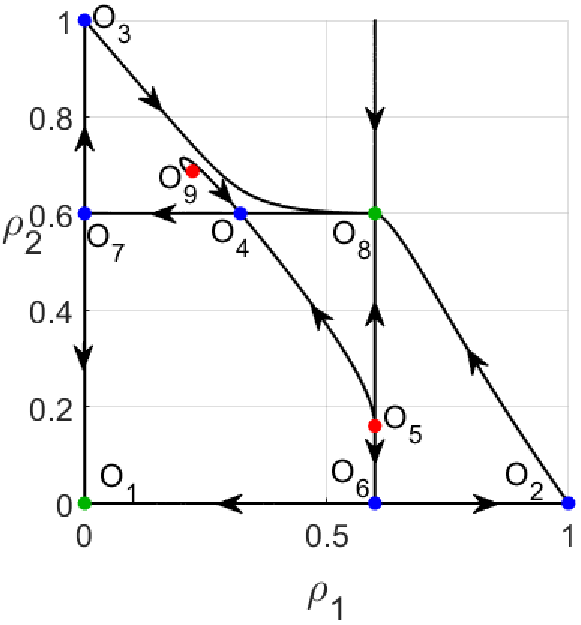}
    }\\
    \subfigure[Region P]
    {
        \includegraphics[width = .23\linewidth]{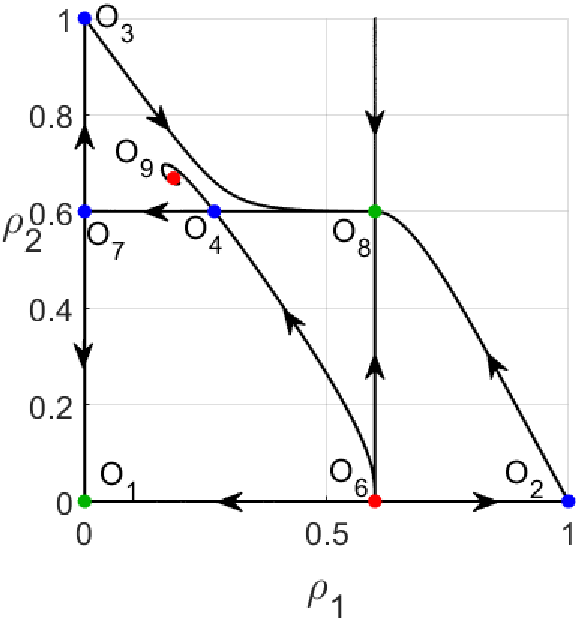}
    }
    \subfigure[Region Q]
    {
        \includegraphics[width = .23\linewidth]{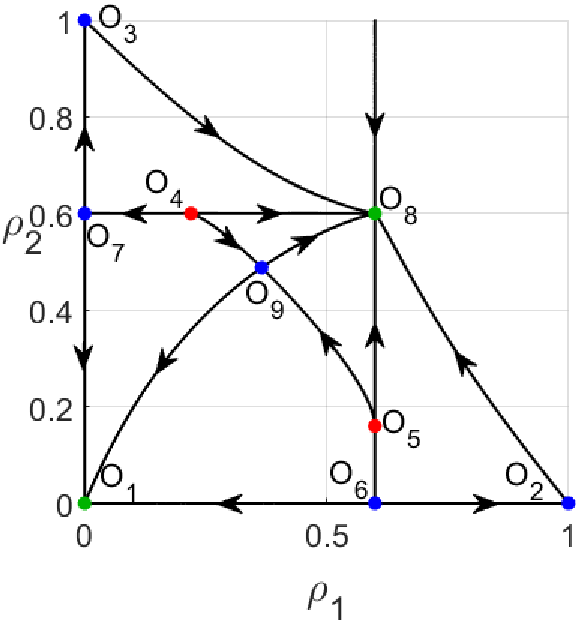}
    }
    \subfigure[Region R]
    {
        \includegraphics[width = .23\linewidth]{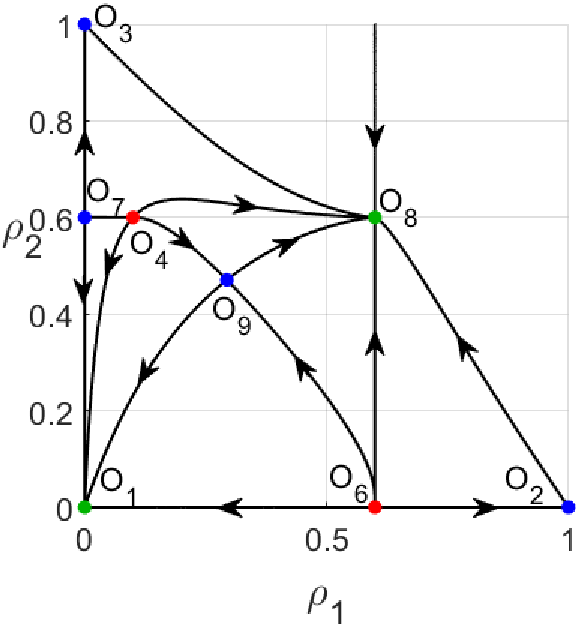}
    }
    \subfigure[Region S]
    {
        \includegraphics[width = .23\linewidth]{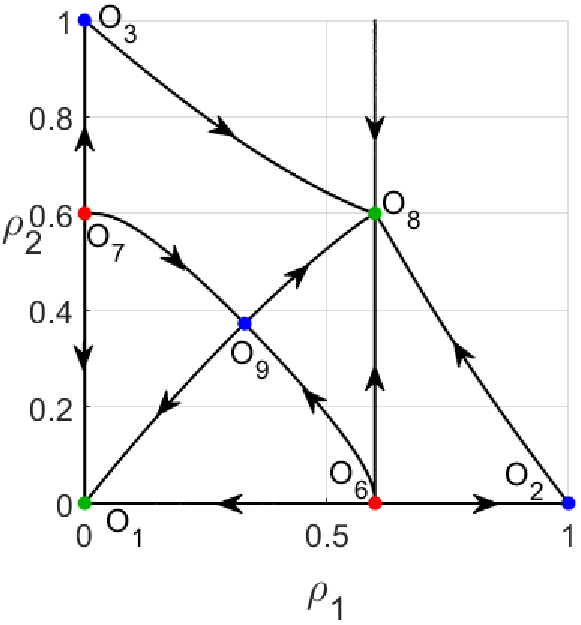}
    }
    \caption{Phase portraits in various regions of Fig.~\ref{Bif_diag_syst2_2d}.}
    \label{PP_syst2_2d}
\end{figure}

Phase portraits of Eqs.~\eqref{LV_system2} for all marked areas are shown in Fig.~\ref{PP_syst2_2d}. Similarly with the first system, the study of two-dimensional Eqs.~\eqref{LV_system2_2d} should provide information about the dynamics of a three-dimensional system, in which the main attention will be paid again to the existence of a heteroclinic cycle. The trajectories of two-dimensional systems on invariant planes $\rho_i=0$ will be components of a three-dimensional heteroclinic cycle. As for the first system on invariant planes $\rho_i=0$, closed trajectories including an unstable node, saddle equilibrium states and a stable node at the origin are of interest. Moreover, the unstable node and one of two saddles should lie on different coordinate axes. Among all other regions on the $\alpha\beta-plane$ $F$, $G$ and $L$ meet all the requirements. In these regions, there are trajectories connecting unstable nodes, saddle equilibrium states and a stable node $O_1$. For example, in Fig.~\ref{PP_syst2_2d} (e), for the regions $F$ and $G$, there is a contour containing the equilibrium states $O_6$, $O_8$, $O_7$ and $O_1$. This contour will be an integral part of the heteroclinic cycle in the three-dimensional phase space of the initial system.

\subsection{A study of Eqs.~\eqref{LV_system2}}
The equilibria of Eqs.~\eqref{LV_system2} and their eigenvalues are given in the Table~\ref{table_6}.

\begin{table}[H]
    \centering
    \begin{tabular}{| c | c | c | c | c |}
        \hline
        № & Equilibrium state & \multicolumn{3}{c|}{Eigenvalues}\\
        \hline
        $1$ & $(0, 0, 0)$ & $-\gamma$ & $-\gamma$ & $-\gamma$\\
        \hline
        $2$ & $(1, 0, 0)$ & $\gamma - 1$ & $\gamma(\alpha - 1)$ & $\gamma(\beta - 1)$\\
        \hline
        $3$ & $(0, 1, 0)$ & $\gamma - 1$ & $\gamma(\alpha - 1)$ & $\gamma(\beta - 1)$\\
        \hline
        $4$ & $(0, 0, 1)$ & $\gamma - 1$ & $\gamma(\alpha - 1)$ & $\gamma(\beta - 1)$\\
        \hline
        $5$ & $(\gamma, 0, 0)$ & $\gamma(\alpha \gamma - 1)$ & $\gamma(\beta \gamma - 1)$ & $\gamma(1 - \gamma)$\\
        \hline
        $6$ & $(0, \gamma, 0)$ & $\gamma(\alpha \gamma - 1)$ & $\gamma(\beta \gamma - 1)$ & $\gamma(1 - \gamma)$\\
        \hline
        $7$ & $(0, 0, \gamma)$ & $\gamma(\alpha \gamma - 1)$ & $\gamma(\beta \gamma - 1)$ & $\gamma(1 - \gamma)$\\
        \hline
        $8$ & $(\gamma, \gamma, 0)$ & $\gamma(\alpha \gamma + \beta \gamma - 1)$ & $\gamma(1 - \gamma - \alpha \gamma)$ & $\gamma(1 - \gamma - \beta \gamma)$\\
        \hline
        $9$ & $(\gamma, 0, \gamma)$ & $\gamma(\alpha \gamma + \beta \gamma - 1)$ & $\gamma(1 - \gamma - \alpha \gamma)$ & $\gamma(1 - \gamma - \beta \gamma)$\\
        \hline
        $10$ & $(0, \gamma, \gamma)$ & $\gamma(\alpha \gamma + \beta \gamma - 1)$ & $\gamma(1 - \gamma - \alpha \gamma)$ & $\gamma(1 - \gamma - \beta \gamma)$\\
        \hline
        $11$ & $(\gamma, \gamma, \gamma)$ & $\gamma(1 - \gamma - \alpha \gamma - \beta \gamma)$ & $\gamma(1 - \gamma - \alpha \gamma - \beta \gamma)$ & $\gamma(1 - \gamma - \alpha \gamma - \beta \gamma)$\\
        \hline
        $12$ & $(1 - \alpha \gamma, \gamma, 0)$ & $\gamma(\alpha + \beta \gamma - \alpha^2 \gamma - 1)$ & $\gamma(\alpha \beta \gamma - \beta - \gamma + 1)$ & $-(\gamma - \frac{1}{\alpha})(\gamma - \frac{1}{\alpha + 1})$\\
        \hline
        $13$ & $(\gamma, 0, 1 - \alpha \gamma)$ & $\gamma(\alpha + \beta \gamma - \alpha^2 \gamma - 1)$ & $\gamma(\alpha \beta \gamma - \beta - \gamma + 1)$ & $-(\gamma - \frac{1}{\alpha})(\gamma - \frac{1}{\alpha + 1})$\\
        \hline
        $14$ & $(0, 1 - \alpha \gamma, \gamma)$ & $\gamma(\alpha + \beta \gamma - \alpha^2 \gamma - 1)$ & $\gamma(\alpha \beta \gamma - \beta - \gamma + 1)$ & $-(\gamma - \frac{1}{\alpha})(\gamma - \frac{1}{\alpha + 1})$\\
        \hline
        $15$ & $(\gamma, 1 - \beta \gamma, 0)$ & $\gamma(\beta + \alpha \gamma - \beta^2 \gamma - 1)$ & $\gamma(\alpha \beta \gamma - \alpha - \gamma + 1)$ & $-(\gamma - \frac{1}{\beta})(\gamma - \frac{1}{\beta + 1})$\\
        \hline
        $16$ & $(1 - \beta \gamma, 0, \gamma)$ & $\gamma(\beta + \alpha \gamma - \beta^2 \gamma - 1)$ & $\gamma(\alpha \beta \gamma - \alpha - \gamma + 1)$ & $-(\gamma - \frac{1}{\beta})(\gamma - \frac{1}{\beta + 1})$\\
        \hline
        $17$ & $(0, \gamma, 1 - \beta \gamma)$ & $\gamma(\beta + \alpha \gamma - \beta^2 \gamma - 1)$ & $\gamma(\alpha \beta \gamma - \alpha - \gamma + 1)$ & $-(\gamma - \frac{1}{\beta})(\gamma - \frac{1}{\beta + 1})$\\
        \hline
        $18$ & $(\frac{\alpha + \beta \gamma - \alpha^2 \gamma - 1}{\alpha \beta - 1}, \frac{\beta + \alpha \gamma - \beta^2 \gamma - 1}{\alpha \beta - 1}, \gamma)$ & $\lambda_1$ & $\lambda_2$ & $\lambda_3$\\
        \hline
        $19$ & $(\frac{\beta + \alpha \gamma - \beta^2 \gamma - 1}{\alpha \beta - 1}, \gamma, \frac{\alpha + \beta \gamma - \alpha^2 \gamma - 1}{\alpha \beta - 1})$ & $\lambda_1$ & $\lambda_2$ & $\lambda_3$\\
        \hline
        $20$ & $(\gamma, \frac{\alpha + \beta \gamma - \alpha^2 \gamma - 1}{\alpha \beta - 1}, \frac{\beta + \alpha \gamma - \beta^2 \gamma - 1}{\alpha \beta - 1})$ & $\lambda_1$ & $\lambda_2$ & $\lambda_3$\\
        \hline
        $21$ & $(\frac{\alpha - 1}{\alpha \beta - 1}, \frac{\beta - 1}{\alpha \beta - 1}, 0)$ & $\mu_1$ & $\mu_2$ & $\mu_3$\\
        \hline
        $22$ & $(\frac{\beta - 1}{\alpha \beta - 1}, 0, \frac{\alpha - 1}{\alpha \beta - 1})$ & $\mu_1$ & $\mu_2$ & $\mu_3$\\
        \hline
        $23$ & $(0, \frac{\alpha - 1}{\alpha \beta - 1}, \frac{\beta - 1}{\alpha \beta - 1})$ & $\mu_1$ & $\mu_2$ & $\mu_3$\\
        \hline
        $24$ & $(1 - \gamma(\alpha + \beta), \gamma, \gamma)$ & $\nu_1$ & $\nu_2$ & $\nu_3$ \\
        \hline
        $25$ & $(\gamma, 1 - \gamma(\alpha + \beta), \gamma)$ & $\nu_1$ & $\nu_2$ & $\nu_3$\\
        \hline
        $26$ & $(\gamma, \gamma, 1 - \gamma(\alpha + \beta))$ & $\nu_1$ & $\nu_2$ & $\nu_3$\\
        \hline
        $27$ & $(\frac{1}{\alpha + \beta + 1}, \frac{1}{\alpha + \beta + 1}, \frac{1}{\alpha + \beta + 1})$ & $\frac{\gamma(1 + \alpha + \beta) - 1}{\alpha + \beta + 1}$ & ${\scriptstyle \frac{(\gamma(1 + \alpha + \beta) - 1)(\alpha + \beta - 2 + (\beta - \alpha)\sqrt 3 i)}{2(\alpha + \beta + 1)^2}}$ & ${\scriptstyle \frac{(\gamma(1 + \alpha + \beta) - 1)(\alpha + \beta - 2 - (\beta - \alpha)\sqrt 3 i)}{2(\alpha + \beta + 1)^2}}$\\
        \hline
    \end{tabular}
    \caption{Equilibria of Eqs.~\eqref{LV_system2}.}
    \label{table_6}
\end{table}

Expressions for the eigenvalues $\lambda_{1, 2, 3}$, as well as the characteristic equation from which they are determined, are too cumbersome. 
The eigenvalues $\mu_{1, 2, 3}$ are determined from the characteristic equation

$$\scriptstyle \left(\mu^2 + \frac{(\alpha - 1)(\alpha + \gamma - \alpha \beta \gamma - 1) + (\beta - 1)(\beta + \gamma - \alpha \beta \gamma - 1)}{(\alpha \beta - 1)^2}\mu - \frac{(\alpha - 1)(\alpha + \gamma - \alpha \beta \gamma - 1)(\beta - 1)(\beta + \gamma - \alpha \beta \gamma - 1)}{(\alpha \beta - 1)^3}\right)\left(\mu - \frac{\alpha^2 + \beta^2 + \alpha \beta + \alpha + \beta - 1}{\alpha \beta - 1}\right) = 0.$$

The eigenvalues $\nu_{1, 2, 3}$ are determined from the characteristic equation
$${\scriptstyle \left(\nu + (\alpha \gamma + \beta \gamma - 1)(\alpha \gamma + \beta \gamma + \gamma - 1)\right)\left(\nu^2 - \gamma(\alpha \gamma + \beta \gamma + \gamma - 1)(\alpha + \beta - 2)\nu + \gamma^2 (\alpha \gamma + \beta \gamma + \gamma - 1)^2(\alpha - 1)(\beta - 1)\right) = 0.}$$

The subset $0 \leq \rho_1, \rho_2, \rho_3 \leq \gamma$ of the phase space of Eqs.~\eqref{LV_system2} is invariant. In order to simplify the study of the system, we impose conditions on $\alpha, \beta$ parameters under which the equilibria $O_{12}$-$O_{26}$ are not in this subset.

$O_{12}$-$O_{14}$ equilibria have a coordinate $1-\alpha\gamma$ that goes beyond at $\alpha > \frac{1}{\gamma}$ (region S) and at $\alpha < \frac{1}{\gamma} - 1$ (regions $A$-$L$).

$O_{15}$-$O_{17}$ equilibria have a coordinate $1-\beta\gamma$ that goes beyond at $\beta > \frac{1}{\gamma}$ (regions $F$, $G$, $L$, $N$, $P$, $R$ and $S$) and at $\beta < \frac{1}{\gamma} - 1$ (regions $A$ and $B$).

$O_{18}$-$O_{20}$ equilibria have a coordinate $\frac{\alpha + \beta \gamma - \alpha^2 \gamma - 1}{\alpha \beta - 1}$ that goes beyond in the regions $E$, $F$, $G$, $J$-$N$, $P$ and $R$.

$O_{21}$-$O_{23}$ equilibria have a coordinates $\frac{\alpha - 1}{\alpha \beta - 1}$ and $\frac{\beta - 1}{\alpha \beta - 1}$ that go beyond in the regions $A$-$G$ and $I$-$P$.

$O_{24}$-$O_{26}$ equilibria have a coordinate $1 - \gamma(\alpha + \beta)$ that goes beyond at $\beta > \frac{1}{\gamma} - \alpha$ (regions $F$, $G$, $J$-$S$) and at $\beta < \frac{1}{\gamma} - \alpha - 1$ that is never met in the current regions.

\begin{table}[H]
    \centering
    \begin{tabular}{| c | c | c | c |}
        \hline
        Equilibrium state & \multicolumn{3}{c|}{Eigenvalues}\\
        \hline
        $O_1(0, 0, 0)$ & $-\gamma$ & $-\gamma$ & $-\gamma$\\
        \hline
        $O_5(\gamma, 0, 0)$ & $\gamma(\alpha \gamma - 1)$ & $\gamma(\beta \gamma - 1)$ & $\gamma(1 - \gamma)$\\
        \hline
        $O_6(0, \gamma, 0)$ & $\gamma(\alpha \gamma - 1)$ & $\gamma(\beta \gamma - 1)$ & $\gamma(1 - \gamma)$\\
        \hline
        $O_7(0, 0, \gamma)$ & $\gamma(\alpha \gamma - 1)$ & $\gamma(\beta \gamma - 1)$ & $\gamma(1 - \gamma)$\\
        \hline
        $O_8(\gamma, \gamma, 0)$ & $\gamma(\alpha \gamma + \beta \gamma - 1)$ & $\gamma(1 - \gamma - \alpha \gamma)$ & $\gamma(1 - \gamma - \beta \gamma)$\\
        \hline
        $O_9(\gamma, 0, \gamma)$ & $\gamma(\alpha \gamma + \beta \gamma - 1)$ & $\gamma(1 - \gamma - \alpha \gamma)$ & $\gamma(1 - \gamma - \beta \gamma)$\\
        \hline
        $O_{10}(0, \gamma, \gamma)$ & $\gamma(\alpha \gamma + \beta \gamma - 1)$ & $\gamma(1 - \gamma - \alpha \gamma)$ & $\gamma(1 - \gamma - \beta \gamma)$\\
        \hline
        $O_{11}(\gamma, \gamma, \gamma)$ & $\gamma(1 - \gamma - \alpha \gamma - \beta \gamma)$ & $\gamma(1 - \gamma - \alpha \gamma - \beta \gamma)$ & $\gamma(1 - \gamma - \alpha \gamma - \beta \gamma)$\\
        \hline
        $O_{27}(\frac{1}{\alpha + \beta + 1}, \frac{1}{\alpha + \beta + 1}, \frac{1}{\alpha + \beta + 1})$ & $\frac{\gamma(1 + \alpha + \beta) - 1}{\alpha + \beta + 1}$ & ${\scriptstyle \frac{(\gamma(1 + \alpha + \beta) - 1)(\alpha + \beta - 2 + (\beta - \alpha)\sqrt 3 i)}{2(\alpha + \beta + 1)^2}}$ & ${\scriptstyle \frac{(\gamma(1 + \alpha + \beta) - 1)(\alpha + \beta - 2 - (\beta - \alpha)\sqrt 3 i)}{2(\alpha + \beta + 1)^2}}$\\
        \hline
    \end{tabular}
    \caption{Equilibria of Eqs.~\eqref{LV_system2} located in the invariant set $0 \leq \rho_1, \rho_2, \rho_3 \leq 1$.}
    \label{eq_inside_3d}
\end{table}

Table~\ref{eq_inside_3d} contains a list of equilibria and the corresponding eigenvalues located in the invariant set $0 \leq \rho_1, \rho_2, \rho_3 \leq 1$. It is these equilibrium states that will be considered in the study of switching activity associated with the presence of heteroclinic cycles $G$. Unstable heteroclinic cycles will contain equilibria located on the planes $\rho_i = 0$. The attractors are the equilibrium states $O_1$ -- all neurons are deactivated and $O_{27}$ -- all neurons are active. Depending on the amplitude of the external stimuli, the neural ensemble, which was initially in an unexcited state, will resume its initial state or all the neurons of the ensemble will become active. Transients to one state or another depend on the amplitude of the external stimuli. The boundary of the regions of attraction of stable equilibrium states $O_1$ and $O_{11}$ is an invariant manifold $M$ containing equilibrium states $O_5$, $O_8$, $O_6$, $O_{10}$, $O_7$, $O_9$ and $O_{27}$. This manifold intersects $\rho_i$ planes along the trajectories belonging to the heteroclinic cycle $G$. Switching sequential neural activity is most pronounced in cases when an external stimuli throws the starting point into a small neighborhood of the manifold $M$. It should be noted that heteroclinic cycles include six (and not three as for the first system) saddle states of equilibrium.

\subsubsection{Sequential activity}
\begin{figure}[H]
    \centering
    \subfigure[]
    {
        \includegraphics[width = .4\linewidth]{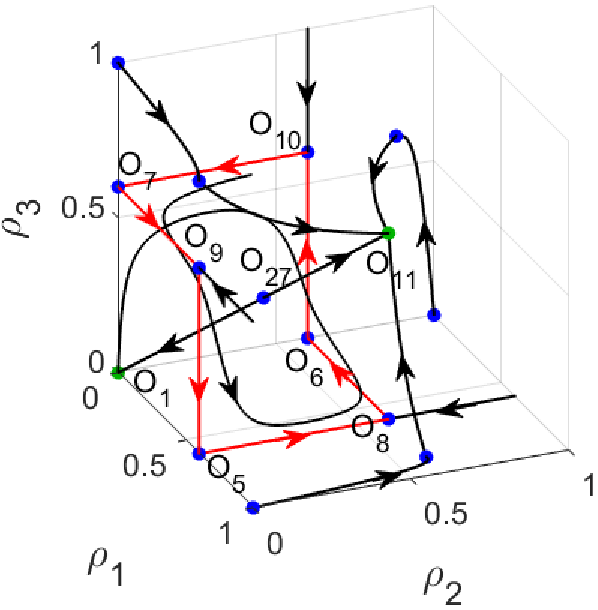}
    }
    \subfigure[]
    {
        \includegraphics[width = .5\linewidth]{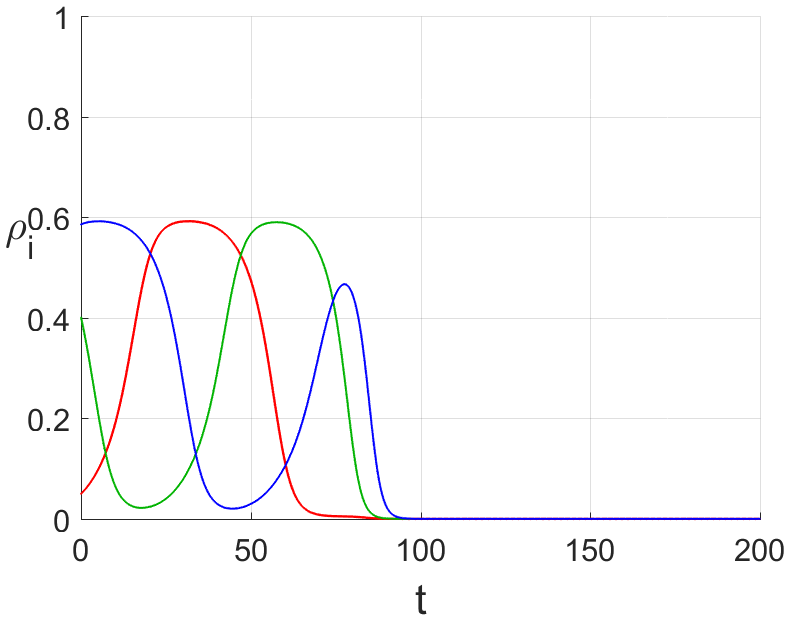}
    }\\
    \subfigure[]
    {
        \includegraphics[width = .4\linewidth]{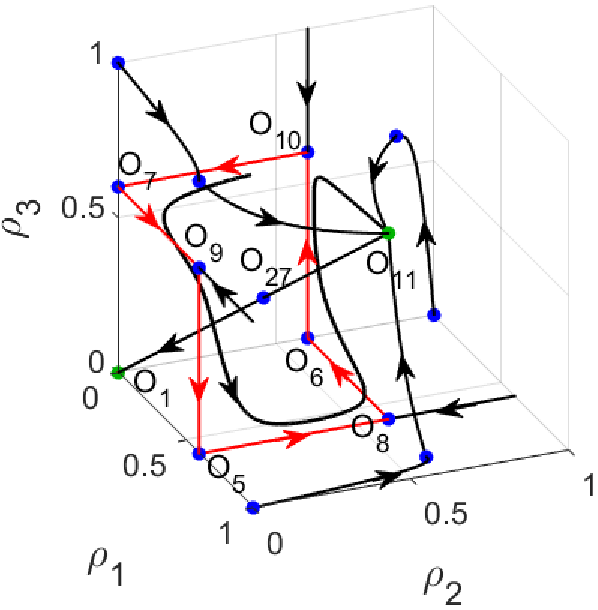}
    }
    \subfigure[]
    {
        \includegraphics[width = .5\linewidth]{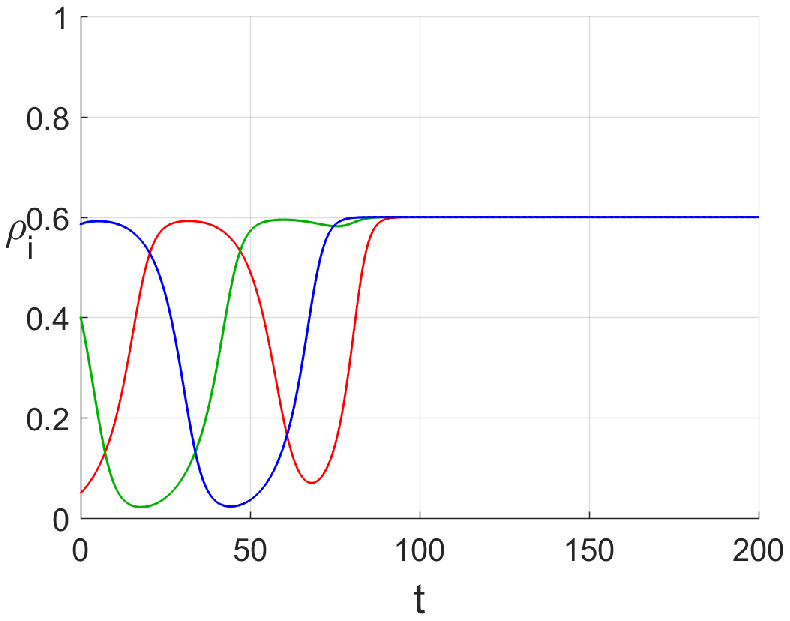}
    }
    \caption{Sequential neural activity. The heteroclinic cycle is marked in red (in Figures (a) and (b)) here and further. Depending on the amplitude of the external stimuli, after each of the elements is activated sequentially once, either deactivation occurs (Figure (a) and (b)), or activation of all elements (Figure (c) and (d)). Parameter values: $\alpha = 0.2$, $\beta = 1.9$, $\gamma = 0.6$. (a), (b) $a = 0.05$, $b = 0.4$, $c = 0.5857$. (c), (d) $a = 0.05$, $b = 0.4$, $c = 0.5858$.}
    \label{Seq_act_2ens_1}
\end{figure}

\begin{figure}[H]
    \centering
    \subfigure[]
    {
        \includegraphics[width = .4\linewidth]{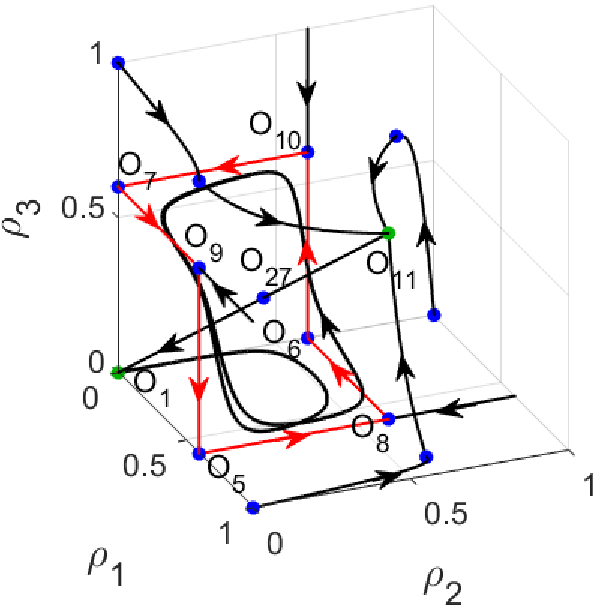}
    }
    \subfigure[]
    {
        \includegraphics[width = .5\linewidth]{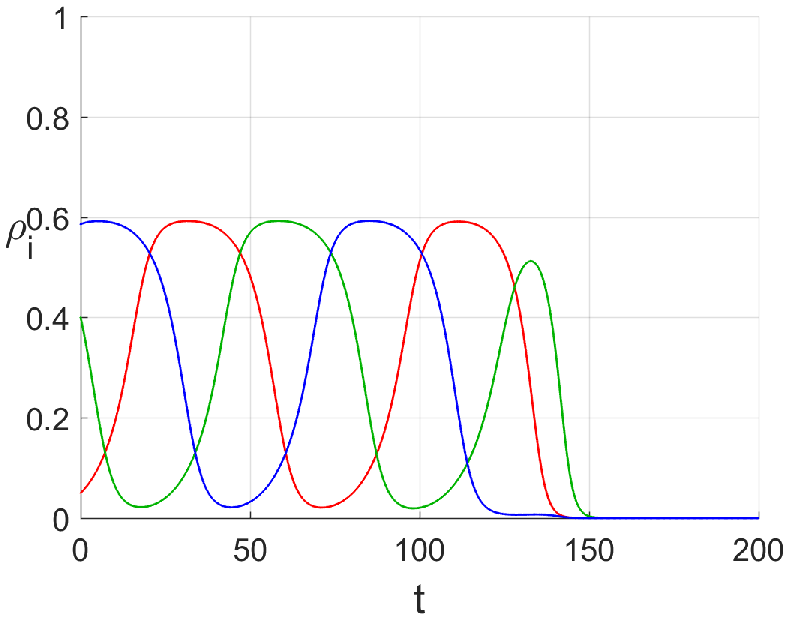}
    }\\
    \subfigure[]
    {
        \includegraphics[width = .4\linewidth]{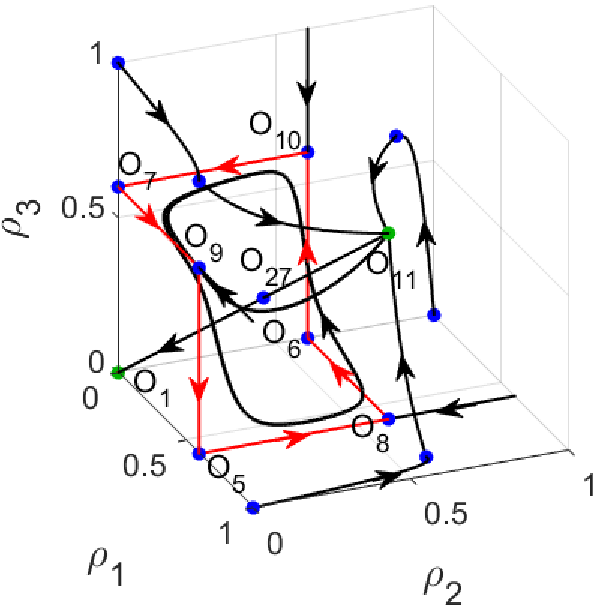}
    }
    \subfigure[]
    {
        \includegraphics[width = .5\linewidth]{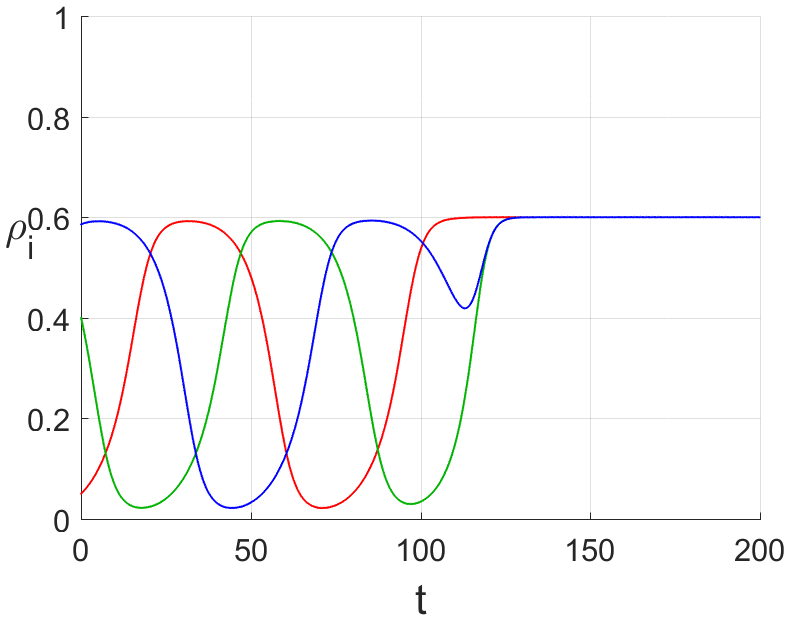}
    }
    \caption{Sequential neural activity. Depending on the amplitude of the external stimuli, after each of the elements is activated sequentially twice, either deactivation occurs (Figure (a) and (b)), or activation of all elements (Figure (c) and (d)). Parameter values: $\alpha = 0.2$, $\beta = 1.9$, $\gamma = 0.6$. (a), (b) $a = 0.05$, $b = 0.4$, $c = 0.585745$. (c), (d) $a = 0.05$, $b = 0.4$, $c = 0.585746$.}
    \label{Seq_act_2ens_2}
\end{figure}

Three-dimensional system described by Eqs.~\eqref{LV_system2}, like Eqs.~\eqref{LV_system1}, can exhibit damping sequential activity. Figures.~\ref{Seq_act_2ens_1} and \ref{Seq_act_2ens_2} show examples of such activity. Depending on the external stimuli, the following dynamics can be obtained:

\begin{itemize}\itemsep=-2pt
    \item sequential one-time excitation of all three neurons (Fig.~\ref{Seq_act_2ens_1} with the subsequent transition of all neurons into an inactive state defined by the equilibrium state $O_1$ ((a) and (b)) or into the active state determined by the equilibrium state $O_{11}$ ((c) and (d)). 

    \item sequential multiple excitation of all three neurons (Fig.~\ref{Seq_act_2ens_2} with subsequent damping ((a) and (b)) or with the transition to the active state of all neurons ((c) and (d)). 
\end{itemize}

In addition, as for the first system, activation of one or two neurons is possible.

\section{Summary and conclusions}

In this paper

\begin{itemize}\itemsep=-2pt
    \item Two dynamical systems based on the generalized Lotka-Volterra were proposed. Ensembles consist of excitable elements with excitatory couplings.

    \item The existence of heteroclinic cycles in the phase space of models is shown. The presence of a heteroclinic cycle in the phase space makes it possible to simulate the effect of winnerless competition between modes – sequential switching of ensemble activity between individual neurons or groups of neurons.

    \item In the proposed models, heteroclinic cycles are not stable, which leads to a complete damping of the ensemble's activity after some time.

    \item Depending on the external stimuli, one, two or three neurons can be activated.

    \item The second model, depending on the amplitude of the external stimuli, can demonstrate a damping sequential neural activity with an exit to an inactive state, or to a permanent active state.

\end{itemize}

The proposed models will be used to construct networks of neural elements and to study their dynamics.

This paper is devoted to Prof. V. N. Belykh, who is outstanding scientist in  the  field of nonlinear dynamics.

\section*{Appendix. Relationship of systems~\eqref{LV_system1} and \eqref{LV_system2} with the population dynamics equation}
\setcounter{equation}{0}

When intraspecific competition is taken into account , the dynamics of population size can be described by the following equation \cite{bazykin1998nonlinear}
\begin{equation} \label{intraspecies_competition}
\dot x = x \left(\frac{b_0 x}{N + x} - d_0 - e x \right),
\end{equation}
where $x$ is the population size, $b_0 x$ is the specific rate of reproduction (fertility), $d_0$ is the specific rate of death of individuals (mortality), $e$ is the coefficient characterizing the influence of competition on fertility and mortality, $N$ is the population density at which the average time between consecutive contacts of one the individual is equal to the minimum time between successive fertilizations of the female.

If we decompose the right side of Eq.~\eqref{intraspecies_competition} into a power series, we get
\begin{equation} \label{intraspecies_competition_series}
\dot x = x \left(-d_0 + \left(\frac{b_0}{N} - e\right)x - \frac{b_0 x^2}{N^2} + \frac{b_0 x^3}{N^3} - \frac{b_0 x^4}{N^4} + \frac{b_0 x^5}{N^5} - ...\right),
\end{equation}
where the coefficient $\frac{b_0}{N} - e > 0$.

If in Eq.~\eqref{intraspecies_competition_series} we restrict ourselves to terms of the second order, then we get the equation
\begin{equation}
\dot x = x \left(-d_0 + \left(\frac{b_0}{N} - e\right)x\right) = d_0 x\left(-1 + \frac{b_0 - e N}{d_0 N} x\right).
\end{equation}
By the substitution $\frac{b_0 - e N}{d_0 N} x = \rho$, we get
\begin{equation}
\frac{d \rho}{d t} = d_0 \rho(-1 + \rho).
\end{equation}
If we now substitute the variable $t$ with $\frac{\tau}{d_0}$, then we get Eq.~\eqref{LV_equation1}.

If the equation \eqref{intraspecies_competition_series} takes into account terms up to and including the third order, then we get the equation
\begin{equation} \label{intraspecies_competition_series_trunc2}
\dot x = x \left(-d_0 + \left(\frac{b_0}{N} - e\right)x - \frac{b_0 x^2}{N^2}\right).
\end{equation}

On condition $\frac{b_0}{N} - e > \frac{2}{N}\sqrt{b_0 d_0}$ polynomial $f(x) = -d_0 + \left(\frac{b_0}{N} - e\right)x - \frac{b_0 x^2}{N^2}$ has two different real roots: $$x_{1,2} = \frac{\frac{b_0}{N} - e \pm \sqrt D}{\frac{2 b_0}{N^2}},$$ where $D = \left(\frac{b_0}{N} - e\right)^2 - \frac{4 b_0 d_0}{N^2}$ is the discriminant of the polynomial $f(x)$. That is $x_{1} > x_{2} > 0$. Then Eq.~\eqref{intraspecies_competition_series_trunc2} can be written as
\begin{equation}
\dot x = \frac{b_0}{N^2}x(x_2 - x)(x - x_1) = \frac{b_0}{N^2} x_1^2 x \left(\frac{x_2}{x_1} - \frac{x}{x_1}\right)\left(\frac{x}{x_1} - 1 \right).
\end{equation}
By the substitution $\frac{x}{x_1} = \rho$, we get
\begin{equation}
x_1 \frac{d \rho}{d t} = \frac{b_0}{N^2} x_1^3 \rho \left(\frac{x_2}{x_1} - \rho \right)(\rho - 1).
\end{equation}
By the substitution $\frac{x_1^2 b_0}{N^2} t = \tau$, $\frac{x_2}{x_1} = \gamma$ we get  Eq.~\eqref{LV_equation2}.

\section*{Funding}
The results in Section 2  were supported by the Ministry of Science and Higher Education of Russian Federation (Project No. 0729-2020-0036), the results in Section 3 were supported by RSF grant \# 23-12-00180.


\end{document}